\def\PN#1{\uuline{#1}}

\def\diff{\mathrm{d}}

\def\N{\mathcal{N}}
\def\F{\mathcal{F}}

\def\n{\mathrm{n}}
\def\p{\mathrm{p}}
\def\q{\mathrm{q}}
\def\s{\mathrm{s}}
\def\v{\mathrm{v}}

\def\A{\mathrm{A}}
\def\B{\mathrm{B}}
\def\C{\mathrm{C}}
\def\D{\mathrm{D}}

\def\exp{\mathrm{exp}}

\def\vecr{\mathbf{r}}
\def\vecp{\mathbf{p}}
\def\veck{\mathbf{k}}
\def\vecn{\mathbf{n}}
\def\Ntest{N_{\textrm{test}}}

\def\eF{\epsilon_{\textrm{F}}}
\def\vF{v_{\textrm{F}}}

\def\bbsty#1#2#3{{\bf #1}, #2 (#3)}	% APS template
\documentclass[prc,aps,twocolumn,psfig,preprintnumbers,amsmath,amssymb]{revtex4-1}
\usepackage{graphicx}\usepackage{epsfig}%\usepackage{portland}
\usepackage{amsmath}\usepackage{amssymb}\usepackage{amsfonts}
\usepackage{bm}
\usepackage{exscale}%\usepackage{rotating}
\usepackage{dcolumn}    % Align table columns on decimal point
\usepackage{color}
\usepackage{soul}
\usepackage{ulem}
\begin{document}
%\begin{frontmatter}
%\preprint{preprintnumber}
%
%   --- TITLE
%
\title{Inhomogeneity growth in two-component fermionic systems}
%
%   --- AUTHORS
%
\author{P.Napolitani$^1$}
\author{M.Colonna$^2$}
%
%   --- AFFILIATIONS
%
\affiliation{	$^1$ IPN, CNRS/IN2P3, Universit\'e Paris-Sud 11, Universit\'e Paris-Saclay, 91406 Orsay Cedex, France
		\\		$^2$ INFN-LNS, Laboratori Nazionali del Sud, 95123 Catania, Italy}
%
%   --- ABSTRACT
%
\begin{abstract}

%*** 3rd version - MC

The dynamics of fermionic many-body systems is investigated in the framework of 
Boltzmann-Langevin (BL) stochastic one-body approaches. 
Within the recently introduced BLOB model,  we examine the interplay between mean-field effects and 
two-body correlations, of stochastic nature, for nuclear matter at moderate temperature and in 
several density conditions, corresponding to stable or mechanically unstable situations. 
Numerical results are compared to analytic expectations for the fluctuation amplitude of isoscalar 
and isovector densities, % which reflect 
probing the link to the properties of the employed effective interaction,
namely symmetry energy (for isovector modes) and incompressibility (for isoscalar modes).
For unstable systems, clusterization is observed. The associated features are compared to analytical 
results for the typical length and time scales characterizing the growth of unstable modes in 
nuclear matter and for the isotopic variance of the emerging fragments.  
We show that the BLOB model is generally better suited than simplified approaches previously 
introduced to solve the BL equation, and it is therefore more advantageous in applications to 
open systems, like heavy ion collisions.

\end{abstract}

%\end{frontmatter}
%
%   --- PACS
%
%\pacs{26.60.+c, 68.35.Rh, 51.30.+i, 05.50.+q}
%
%\pacs{26.60.+c}{Nuclear matter aspects of neutron stars}
%\pacs{68.35.Rh}{Phase transitions and critical phenomena}
%\pacs{51.30.+i}{Thermodynamic properties, equations of state}
%\pacs{05.50.+q}{Lattice theory and statistics (Ising, Potts, etc.)}
%
% 26.60.+c Nuclear matter aspects of neutron stars
% 64.60.-i General studies of phase transitions
% 05.70.Jk Critical point phenomena
% 68.35.Rh Phase transitions and critical phenomena
% 26. Nuclear astrophysics
% 05.70.Fh Phase transitions: general studies
% 05.50.+q Lattice theory and statistics (Ising, Potts, etc.)
% 05.70.-a Thermodynamics
% 51.30.+i Thermodynamic properties, equations of state
%
\maketitle
%
%
%%%%%%%%%%%%%%%%%%%%%%%%%%%%%%%%%%%%%%%%%%%%%%%%%%%%%%%%%%%%%%%%%%%
%
%       T E X T
%
%%%%%%%%%%%%%%%%%%%%%%%%%%%%%%%%%%%%%%%%%%%%%%%%%%%%%%%%%%%%%%%%%%%
%

%%%%%     %%  %%%  %%%     %%%     %%      %%      %%     %%%%%%%%%
%%%%%  %%%%%  %%%  %%  %%%  %%  %%  %%%  %%%%  %%%%%%  %%  %%%%%%%%
%%%%%  %%%%%       %%  %%%  %%  %%  %%%  %%%%    %%%%  %%  %%%%%%%%
%%%%%  %%%%%  %%%  %%       %%     %%%%  %%%%  %%%%%%     %%%%%%%%%
%%%%%     %%  %%%  %%  %%%  %%  %%%%%%%  %%%%      %%  %%  %%%%%%%%
\section{Introduction}

%phenomenology: occurrence of violent large amplitude...

%{\it The dynamics of many-body interacting systems has always been an exciting and attractive field in different domains of physics.}
The dynamics of many-body interacting systems is a long-standing investigation embracing various domains at the boundary between collective and chaotic processes.
%	In the nuclear context, the splitting of a composite system into fragments under violent perturbations signs the occurrence of the most catastrophic effect produced by large-amplitude fluctuations of the neutron and proton content.}
%	Such phenomenology, which is common in heavy-ion collisions at Fermi energies~\cite{EPJAtopicalWCI2006,EPJAtopicalNSE2014}, also characterises other fields, like solid-state physics (examples are metal clusters~\cite{Calvayrac2000,Fennel2010} or electrons in nanosystems~\cite{Chen2005}), ultracold atomic gases~\cite{Dalibard1998,Bloch2008} or some areas of astrophysics~\cite{Horowitz2006,Sebille2011,Schneider2013,Burrello2016}. 
%\PN{and relativistic heavy-ion collisions~\cite{Randrup2012}.}
%{\it The description of the fragmentation process can only be afforded within approaches beyond the 
%mean-field approximation, incorporating the effect of many-body correlations, which induce
%fluctuations in the evolution of the one-body density. } 
%%	We focus thereafter on crucial modelling issues studied in nuclear matter, looking
%%in particular at the dynamics of fluctuations, both in stable systems and in unstable
%%conditions, leading to the disassembly of the system.} 

%\subsection{Context: one-body framework}
%
	From a one-body modelling perspective, the dynamics %collective motion 
of fermionic systems~\cite{Ring1980,Maruhn2010} is efficiently described with the time-dependent Hartree-Fock (TDHF) framework, or time-dependent local density approximation (TDLDA), in condense-matter applications~\cite{Yabana1999,Reinhard2004}, as far as the variance of the involved observables is small and can be neglected.
	If this is not the case, additional beyond mean-field correlations should be included, depending on the degree of excitation~\cite{Abe1996,Chomaz2004,Simenel2010}.
	At low energy, as far as the system can be described in the small-amplitude limit, 
a scheme of coherent-state propagation within the time-dependent generator coordinate method (TDGCM)~\cite{Reinhard1987,Goutte2005}, or a variational approach \`a la Balian-V\'en\' eroni~\cite{Balian1981,Simenel2012} is well suited.
	On the contrary, when the system experiences a violent dynamics, large fluctuations would spontaneously drive the system far away from the one-body TDHF evolution along many different directions, thus determining the shortcoming of the TDHF approximation  (and the above mentioned extensions).
	To address large-amplitude regimes, solutions beyond the single-particle picture may be needed, 
%already in the low-energy regime, 
for instance by propagating non-correlated states~\cite{Lacroix2015}.
	When even the low-energy regime is exceeded, dissipative behaviour results also from in-medium collisions, which are no more hindered by Pauli blocking. 

%	Accordingly, especially 
In presence of mean-field instabilities, the collective dynamics may be driven to a chaotic regime; in a nuclear system, this would result in a highly non-linear process, leading to clusterisation from one-body density fluctuations, and oscillation of the neutron and proton fraction.
	The splitting of a composite system into fragments under violent perturbations signs the occurrence of the most catastrophic effect produced by large-amplitude fluctuations of the neutron and proton content.
	Such phenomenology, which is common in heavy-ion collisions at Fermi energies~\cite{EPJAtopicalWCI2006,EPJAtopicalNSE2014}, also characterises other fields, like solid-state physics (examples are metal clusters~\cite{Calvayrac2000,Fennel2010} or electrons in nanosystems~\cite{Chen2005}), ultracold atomic gases~\cite{Dalibard1998,Bloch2008} or some areas of astrophysics~\cite{Horowitz2006,Sebille2011,Schneider2013,Burrello2016}. 
%\PN{and relativistic heavy-ion collisions~\cite{Randrup2012}.}
The description of the fragmentation process can only be afforded within approaches beyond the 
mean-field approximation, incorporating the effect of many-body correlations, which induce
fluctuations in the evolution of the one-body density.
%	Approaches in the spirit of the extended TDHF (ETDHF) framework~\cite{Wong1978,Wong1979,Lacroix2004}, can efficiently describe the widening of some observable spread widths related to the dissipative character of the process, but they are not adapted to also follow possible bifurcation paths which deviate from the mean trajectory.
	Adapted to such situation, stochastic approaches typically propagate a bunch of mean-field trajectories within various orders of approximations, like stochastic TDHF (STDHF) formulations~\cite{Reinhard1992,Suraud2014,Slama2015,Lacombe2016}, or analogue semiclassical schemes within the Boltzmann-Langevin (BL) transport equation~\cite{Ayik1988,Reinhard1992bis}.

%
%\subsection{Purpose: quantifying fluctuation amplitudes}
%
	In the following we exploit the last mentioned BL approach in the form of the recently introduced 
Boltzmann-Langevin one body (BLOB) model~\cite{Napolitani2013,Rizzo2008} to undertake an exhaustive analysis of the interplay between mean-field and many-body correlations in %periodic 
nuclear matter.
	We focus thereafter on crucial modelling issues, looking
in particular at the dynamics of fluctuations, both in stable systems and in unstable
conditions, leading to the disassembly of the system. 
	The purpose is to examine virtues and limits of the BLOB approach, where the BL equation is solved in full phase space, and of corresponding approximations by carrying out a quantitative study of fluctuation amplitudes, 
and comparing with some analytic expectations which characterise Fermi liquids.
	This analysis is important also in the spirit 
of preparing reliable applications to heavy-ion collisions.
Indeed a good reproduction of the fluctuation dynamics is crucial for the predictions of features, 
such as size and isotopic variances, of the products formed in nuclear reactions and to probe their link
to the properties of the nuclear effective interaction.
%it is more instructive to carry out such study in nuclear matter and not, straightforward, in its final application terrain, i.e. open systems.

	In Sec.~\ref{sec_correlations} we survey some basic steps leading from a stochastic beyond-mean-field framework to the BLOB method, and related approximations.
	In Sec.~\ref{sec_BLNM} 
the dynamics of fluctuations of the one-body density, as given by the linearized BL equation,
is discussed for nuclear matter initialized at moderate temperature and in several density conditions.
Owing to the presence of two components (neutrons and protons), one observes isovector fluctuation modes, 
where  neutrons and protons oscillate out of phase, and isoscalar modes, with neutrons and protons moving
together. 
Isovector fluctuations are of stable nature, reflecting the properties of the nuclear effective interaction
in the isovector channel. The performance of the BLOB model in reproducing analytic expectations for the
isovector variance and, in particular, its link to the nuclear symmetry energy is discussed in 
Sec.~\ref{sec_ivfluctuations}.    
For nuclear matter at suitable density and temperature conditions, isoscalar fluctuations may become
unstable, yielding a growth of the (isoscalar) fluctuation variance, which triggers 
a process of clusterisation. 	Such situation is tested in Sec.~\ref{sec_isfluctuations}.
%two test situations are introduced.
%	First. When neutrons and protons oscillate out of phase, the (isovector) fluctuation variance probes the strength of the symmetry energy. 
%	Such situation will be tested in Sec.~\ref{sec_ivfluctuations}.
%	Second. When neutrons and protons oscillate in phase, fluctuation seeds can yield a growth of the (isoscalar) fluctuation variance in stable and unstable conditions, triggering a process of clusterisation.
%	Such situation will be tested in Sec.~\ref{sec_isfluctuations}.
	In the spirit of connecting nuclear matter to open systems, sec.~\ref{sec_open} explores fluctuation observables related to blobs of matter, which correspond to emerging fragments in open systems.
	Conclusive statements from reviewing the results form Sec.~\ref{sec_conclusion}. 
%
% illustrates the transition from nuclear matter to a corresponding open system, where, in presence of clusterisation, the relation investigated in the first situation determines how the isospin content is expected to be distributed among more or less dense portions of the system, while the second situation explores the possible arising of fragments in stable and unstable conditions.
%	We conclude in Sec.~\ref{sec_conclusion} by confirming that a complete solution of the BL equation in full phase space is more efficient than approximated methods in describing large-amplitude dynamic processes under the constraint of preserving the Fermi statistics. This result is corroborated by the quantitative study on fluctuation observables carried on in this work and it is complemented by a comprehensive discussion on the interplay between mean-field and many-body correlations.

%	To achieve this purpose, since early investigations, it was evident that fluctuations could be handled naturally by describing the propagation of several mean-field trajectories %samples} simultaneously, along a bundle of different dynamical paths.

%%%%%     %%  %%%  %%%     %%%     %%      %%      %%     %%%%%%%%%
%%%%%  %%%%%  %%%  %%  %%%  %%  %%  %%%  %%%%  %%%%%%  %%  %%%%%%%%
%%%%%  %%%%%       %%  %%%  %%  %%  %%%  %%%%    %%%%  %%  %%%%%%%%
%%%%%  %%%%%  %%%  %%       %%     %%%%  %%%%  %%%%%%     %%%%%%%%%
%%%%%     %%  %%%  %%  %%%  %%  %%%%%%%  %%%%      %%  %%  %%%%%%%%
\section{Theoretical survey \label{sec_correlations}}
%\section{Handling N-body correlations \label{sec_correlations}}
%\section{Handling fluctuations from collisional correlations}
%\section{dissipative regime through a stochastic framework}

\subsection{N-body correlations in a stochastic one-body framework in dissipative regimes \label{sec_Nbody}}
It is usual to describe the evolution of an $N$-body system by replacing the Liouville-von Neumann equation
%, which involves $6N$ variables, by 
with the equivalent BBGKY hierarchy which, for a two-body interaction $V_{ij}$, yields the following chain of coupled equations
\begin{eqnarray}
i\hbar\frac{\partial\rho_1   }{\partial t} &=& [k_1           ,\rho_1   ] + \textrm{Tr}_2 [V_{12}       ,\rho_{12}] \notag\\
i\hbar\frac{\partial\rho_{12}}{\partial t} &=& [k_1+k_2+V_{12},\rho_{12}] + \textrm{Tr}_3 [V_{13}+V_{23},\rho_{123}] \notag\\
\dots&&\notag\\
i\hbar\frac{\partial\rho_{1^{\dots} k}}{\partial t} &=& \sum_{i=1}^{k}\Big[k_i+\sum_{j<i}^{k}V_{ij},\rho_{1^{\dots} k}\Big]  \notag\\
     &+&\sum_{i=1}^{k}\textrm{Tr}_{k+1}[V_{i k\!+\!1},\rho_{1^{\dots} k\!+\!1}] \notag\\
\dots&&,
	\label{eq:BBGKY}
\end{eqnarray}
where $\textrm{Tr}_{k}$ is a partial trace involving the many-body density matrix $\rho_{1^{\dots} k}$ (compact notation where the order corresponds to the number of indexes) and $k_i$ are kinetic energy operators.
	This is the avenue for constructing beyond-mean-field approximations, obtained through custom truncations of the hierarchy, or by reducing the complexity of the involved contributions at given orders. 
	For instance, the inclusion of interactions beyond two bodies would be necessary to account for additional nuclear-structure features~\cite{Schuck2016}, or cluster correlations, and the explicit inclusion of correlations beyond the order $k=2$ would be necessary to describe high-coupling regimes~\cite{Lacroix2014}. 

	If, on the other hand, a suited stochastic approach is adopted, simplified higher-order contributions can be introduced even though not explicitly implemented.
	Already in a first-order-truncation scheme ($k\!=\!1$) in a low-energy framework, it was found that a stochastic approach can be used to restore all the BBGKY missing orders approximately~\cite{Lacroix2016}, and generate large-amplitude fluctuations; in this case, a coherent ensemble of mean-field states is propagated along different trajectories from an initial stochastic distribution.
	Such scheme is however insufficient for our purpose, which is addressing dissipative regimes.
	In this case, it is necessary to introduce in-medium collisions in a second-order scheme ($k\!=\!2$), explicitly, i.e. the first two lines of the set~(\ref{eq:BBGKY}), from which kinetic equations are obtained~\cite{Balescu1976,Balian1991,Cassing1992,Bonasera1994}.
	If structure effects are neglected, it is then possible to propagate an incoherent ensemble of mean-field states, supplemented by a fluctuating term, in order to obtain a highly non-linear character of the dynamics.
	The stochastic treatment is not 
%introduced in the initial state 
obtained from exploiting a distribution of initial states 
but, progressing from a single initial state. It acts intermittently all along the temporal evolution, producing successive splits of a given mean-field trajectory $\rho_1$ into subensembles $\rho_1^{(n)}$: 
\begin{equation}
	\rho_1 \longrightarrow \{ \rho_1^{(n)}; n=1, \dots, \textrm{subens.} \}   \;.
\end{equation}
	This pattern then repeats for each element of the subensemble $\rho_1^{(n)}$ till eventually yielding trajectories ordered in bifurcating bundles, each one exhibiting a small variance around the mean trajectory of the corresponding envelope.
	In particular, in the time interval between two successive splits, 
when fluctuations are built up, the system propagates keeping the mean trajectory unchanged within each envelope.
	Thus this time, $\tau$, has to be shorter than the time scales associated with the global effect of the collision integral  and with the mean-field propagation.
	
\subsection{Collisional correlations \label{sec_coll_correlations}}
	This stochastic scheme is equivalent to imposing that $\rho_1^{(n)}$ and $\rho_2^{(n)}$, i.e. the probabilities to find two nucleons, 1 and 2, at two configuration points, are not all the time decorrelated, 
%while fluctuations act keeping the mean $\rho^{(m)}$ unchanged.
so that the two-body density matrix $\rho_{12}^{(n)}$ recovers some correlations of the upper orders of the BBGKY sequence in addition to the standard product of independent one-body densities which builds up the mean-field term.
	We can write at a time $t$: %=t_0+\tau$:
\begin{eqnarray}
	&\rho_{12}^{(n)}(t) = \widetilde{\Omega}_{12}\mathcal{A}_{12}(\rho_1^{(n)}(t)\rho_2^{(n)}(t))\widetilde{\Omega}_{12}^{+}+\delta\rho_{12}^{(n)}(t)  \;;\quad&
	\label{eq:fluctuating_term}\\
	&\langle\delta\rho_{12}^{(n)}(t) \rangle_{\tau} = 0 \;;\;&
	\label{eq:fluctuating_mean}\\
	&\langle\delta\rho_{12}^{(n)}(t) \delta\rho_{12}^{(n)}(t) \rangle_{\tau} = \textbf{gain} + \textbf{loss} \;,&
	\label{eq:fluctuating_secmom}
\end{eqnarray}
	where $\widetilde{\Omega}_{12}$ is the M\o ller wave operator~\cite{Reed1979,Suraud1995} describing the diffusion of a particle with respect to another particle in the nuclear medium, related to a diffusion matrix $G_{12}=V_{12}\widetilde{\Omega}_{12}$, which is, in turn, related to the nucleon-nucleon differential cross section $|G_{12}|^2\!\sim\!\textrm{d}\sigma/\textrm{d}\Omega$. 
	In this respect, the first term of the r.h.s. of eq.(\ref{eq:fluctuating_term}) contains collisional correlations, while the second term introduces a fluctuation of vanishing first moment
%, which is related to a fluctuating collision term, introducing a fluctuation 
around the collision integral~\cite{Abe1996}.
	It should be noticed that the average in Eqs.(4,5) refers to an ensemble of one-body trajectories fluctuating, over the
time $\tau$, around a mean trajectory which follows the Boltzmann equation.
	Setting $\delta\rho_{12}^{(n)}(t)=0$ 
%would impose a full decorrelation between $\rho_1^{(n)}$ and $\rho_2^{(n)}$, and 
%removes the effects of correlations beyond the order $k=2$.
would then reduce to the (quantum) Boltzmann kinetic equation, which corresponds to a second-order truncation of the hierarchy without a fluctuation contribution.

	Finally, the description associated with one single mean-field trajectory $\rho_1^{(n)}$ 
%in an interval of time $\sim \tau$ 
yields the following form of the BL equation
%, similar to STDHF \cite{Reinhard1992}, 
containing an average collision contribution $\bar{I}_{\textrm{coll}}^{(n)}$ and a continuous source of fluctuation seeds $\delta I_{\textrm{coll}}^{(n)}$:
\begin{equation}
	i\hbar\frac{\partial\rho_1^{(n)}}{\partial t} \approx [k_1^{(n)}+V_1^{(n)} , \rho_1^{(n)}] 
		+ \bar{I}_{\textrm{coll}}^{(n)} + \delta I_{\textrm{coll}}^{(n)}   \;.
\label{eq:STDHF}
\end{equation}
%	where the residual terms $\bar{I}_{\textrm{coll}}^{(n)}$ and $\delta I_{\textrm{coll}}^{(n)}$ represent an average collision contribution and a continuous source of fluctuation seeds, respectively. 

	It may be noted that Eq.~(\ref{eq:STDHF}) 
is similar to STDHF~\cite{Reinhard1992}, and it
transforms into an extended TDHF (ETDHF) theory~\cite{Wong1978,Wong1979,Lacroix2004} if the fluctuating term $\delta I_{\textrm{coll}}^{(n)}$ is suppressed;
	ETDHF can in fact efficiently describe the behavior %widening 
of some observable %spread widths 
related to dissipative processes, but it can not follow possible bifurcation paths deviating from the mean trajectory.
	Through a Wigner transform we can then replace Eq.~(\ref{eq:STDHF}) by a corresponding set of semiclassical BL trajectories: 
\begin{equation}
	\frac{\partial f^{(n)}}{\partial t} = \{h^{(n)} , f^{(n)}\} 
		+ I_{\textrm{UU}}^{(n)} + \delta I_{\textrm{UU}}^{(n)}   \;,
\label{eq:SMF}
\end{equation}
where the evolution of a statistical ensemble of Slater determinants is replaced by the evolution of an ensemble of distribution functions $f^{(n)}$, which at equilibrium correspond to a Fermi statistics.
	$h^{(n)}$ is the effective Hamiltonian acting on $f^{(n)}$.
	The residual average and fluctuating contributions of Eq.~(\ref{eq:STDHF}) are replaced by Uehling-Uhlenbeck (UU) analogue terms.
	$I_{\textrm{UU}}^{(n)}$ is related to the mean number of transitions 
% between phase-space elementary volumes d$\nu$.
within a single phase-space cell $\Delta V_f$. 
	While conserving single-particle energies, $\delta I_{\textrm{UU}}^{(n)}$ acts as a Markovian contribution expressed through its correlation~\cite{Colonna1994_a} 
\begin{eqnarray}
	&\langle \delta I_{\textrm{UU}}^{(n)}(\vecr,\vecp,t) \delta I_{\textrm{UU}}^{(n)}(\vecr'\!,\vecp'\!,t')\rangle 
	= {\bf gain + loss} =& \notag\\
	&= 2\mathcal{D}(\vecr,\vecp;\vecr'\!,\vecp'\!,t')\delta(t-t') \;,&
\label{eq:correlation}
\end{eqnarray}
where $\mathcal{D}$ is a %space-local  
diffusion coefficient \cite{Ayik1988}.

\subsection{Obtaining the BLOB description: fluctuations in full phase space \label{sec_correlations_BLOB}}
From Eq.(4) %Eq.~(\ref{eq:fluctuating_mean})  
and from the procedure detailed in ref.~\cite{Ayik1990},
we assume %deduce 
that the fluctuating term $\delta I_{\textrm{UU}}^{(n)}$ in Eq.~(\ref{eq:SMF}) should involve the same contributions composing the average collision term $I_{\textrm{UU}}^{(n)}$, i.e. the transition and the Pauli-blocking terms.
	This implies that also $\delta I_{\textrm{UU}}^{(n)}$ should be expressed in terms of one-body distribution functions. 
%treated as a one-body extension of the collision term $I_{\textrm{UU}}^{(n)}$.
This latter possibility can be exploited by replacing the residual terms $(I_{\textrm{UU}}^{(n)} + \delta I_{\textrm{UU}}^{(n)})$ by a similar UU-like term which respects the Fermi statistics both for the occupancy mean value and for the occupancy variance.
	In this case, 
for a free Fermi gas, 
the occupancy variance at equilibrium should be equal to $f^{(n)}(1-f^{(n)})$ in a phase-space cell $h^3$ and correspond to the movement of extended portions of phase space which have the size of a nucleon, i.e. the residual term $(I_{\textrm{UU}}^{(n)} + \delta I_{\textrm{UU}}^{(n)})$ should carry nucleon-nucleon correlations \cite{Bauer1987}.

	A natural solution to satisfy such requirement is to rewrite the residual contribution in the form of a rescaled UU collision term where a single binary collision involves extended phase-space portions of equal isospin $A$, $B$ to simulate wave packets, and Pauli-blocking factors act on the corresponding final states $C$, $D$, also treated as extended phase-space portions.
	The choice of defining each phase-space portion $A$, $B$, $C$ and $D$ so that its isospin content is either $1$ or $-1$ is necessary to preserve the Fermi statistics for both neutrons and protons,
%to produce isovector fluctuations, 
and it imposes that blocking factors are defined accordingly in phase-space cells for the given isospin species.
	The above conditions lead to the BLOB equations~\cite{Napolitani2013}:
\begin{eqnarray}
	\frac{\partial f^{(n)}}{\partial t} - \{h^{(n)} , f^{(n)}\} 
		= I_{\textrm{UU}}^{(n)} + \delta I_{\textrm{UU}}^{(n)}  = \notag\\
	= g\int\frac{\diff\vecp_b}{h^3}\,
	\int
	W({\scriptstyle\A\B\leftrightarrow\C\D})\;
	F({\scriptstyle\A\B\rightarrow\C\D})\;
	\diff\Omega\;,
\label{eq:BLOB}
\end{eqnarray}
where $g$ is the degeneracy factor. 
	$W$ is the transition rate, in terms of relative velocity between the two colliding phase-space portions and differential nucleon-nucleon cross section
\begin{equation}
	W({\scriptstyle\A\B\leftrightarrow\C\D}) = |v_\A\!-\!v_\B| \frac{\diff\sigma}{\diff\Omega}\;.
\label{eq:transition_rate}
\end{equation}
	$F$ contains the products of occupancies and vacancies of initial and final states over their full phase-space extensions.
\begin{equation}
	F({\scriptstyle\A\B\rightarrow\C\D}) =
	\Big[(1\!\!-\!\!{f}_\A)(1\!\!-\!\!{f}_\B) f_\C f_\D - f_\A f_\B (1\!\!-\!\!{f}_\C)(1\!\!-\!\!{f}_\D)\Big]\;.
\label{eq:Pauli_scaled}
\end{equation}
	Details on the implementation of BLOB are given in appendix~\ref{sec_appendix_metrics}.
	In practice, if the test-particle method is employed, so that the system is sampled by $\Ntest$ test-particle per nucleon, phase-space portion $A$, $B$, $C$ and $D$ should be agglomerates of $\Ntest$ test-particles each, and the nucleon-nucleon cross section used in Eq.~(\ref{eq:transition_rate}) should be scaled by the same amount $\Ntest$ (Eq.~(\ref{eq:XSscaling}) ).
	Finally, the stochastic approach exploits the correlations carried in Eq.~(\ref{eq:BLOB}), recovering higher order than the $k\!=\!2$ truncation, and inducing the BL fluctuations-bifurcation scheme.

\subsection{Simplification through the SMF description: fluctuations projected}
%
%	The stochastic term in Eq.~(\ref{eq:SMF}) can be kept separate and treated as a stochastic force related to an external potential $U'$, like in the corresponding semi-classical stochastic mean field (SMF) model~\cite{Colonna1998}. This leads to treatments where fluctuations are implemented only in the coordinate space, i.e. they are projected on a suited coordinate, like the density space, so that $\delta I[f]=\partial_{\vecr}\,U'\;\partial_{\vecp}\,f$.
%	Eq.~(\ref{eq:BLOB}) may be confused with the early approach of Bauer and Bertsch~\cite{Bauer1987}. There is however a very fundamental difference: in Bauer-and-Bertsch approach the Pauli-blocking term is not applied to the involved portions of phase space which are actually interested by the scattering at a given time $t$, but it is applied in average. Such approximation has the effect of loosing the Fermi statistics because the variance $f^{(n)}(1-f^{(n)})$ is not respected~\cite{Chapelle}.
%
%	--- FIGURE collisionvariance
%
\begin{figure}[b!]\begin{center}
	\includegraphics[angle=0, width=.95\columnwidth]{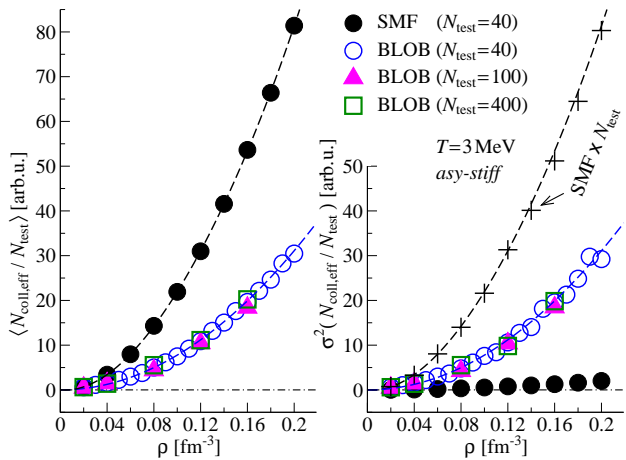}
\end{center}\caption
{
%	Mean-value and variance of the number of effective in-medium collisions per nucleon and per unit time $\delta t=1$ fm/$c$ for SMF and BLOB as a function of density, calculated at equilibrium. %(equilibrium: the average collision rate is constant in time) 
%	Dashed lines, obtained from quadratic fits of the mean-value spectra, are also plotted over the variance spectrum for comparison.
%	The SMF calculation shows that the variance is scaled by a factor equal to $\Ntest$ with respect to the mean.
%	The BLOB calculation shows that the variance equals the mean, regardless of $\Ntest$ values.
%	An asy-stiff parameterisation is used (the characteristics of the interactions are defined in Sec.~\ref{sec_ivfluctuations}).
%
	Mean-value (left) and variance (right) of the number of effective in-medium collisions per nucleon and per unit time $\delta t=1$ fm/$c$ for SMF (full black circles) and BLOB, as a function of density.  The different open symbols correspond to BLOB calculations
with different number of test particles.  
	The dashed-lines on the left panel represent fits to mean-value spectra and they 
are replotted identically in the right panel, for comparison.
The crosses in the righ panel correspond to the SMF variance (black dots) multiplied by $\Ntest$.
%	SMF produces a variance scaled by a factor $\Ntest$ with respect to the mean.
%	BLOB produces a variance equal to the mean, regardless of $\Ntest$ values.
	An asy-stiff parameterisation is used for the symmetry energy (the characteristics of the interactions are defined in Sec.~\ref{sec_ivfluctuations}).
}
\label{fig_collisionvariance}
\end{figure}

	At variance with the above description, the stochastic term in Eq.~(\ref{eq:SMF}) can be kept separate and treated as a stochastic force related to an external potential $U'$, like in the corresponding semi-classical stochastic mean field (SMF) model~\cite{Colonna1998}.
	This leads to treatments where fluctuations are implemented only in the coordinate space, i.e. they are projected on the spacial density.	
	The difference between Eq.~(\ref{eq:BLOB}) and usual stochastic mean-field approaches is that those latter build fluctuations from introducing a well adapted external force or a distribution of initial conditions which should be accurately prepared in advance.
	On the contrary, Eq.~(\ref{eq:BLOB}) introduces fluctuations in full phase space and let them develop spontaneously and continuously over time in a dynamical process.

	Fig.~\ref{fig_collisionvariance} illustrates how this difference affects the statistics of effective in-medium collision (i.e. not Pauli blocked); it compares SMF and BLOB calculations performed in nuclear matter at $3$~MeV temperature and
at different densities in stable conditions, i.e. keeping the system uniform, and at equilibrium, i.e. when the average collision rate is constant in time (these conditions are described in details in Sec.~\ref{sec_ivfluctuations}).
	As a general trend, larger densities provide a larger number of collision candidates, so that, even if also the difficulty in relocating collision partners increases due to Pauli blocking, the resulting number of effective collisions per nucleon grows significantly with density. 
	Despite the use of the same nucleon-nucleon cross section (which produces equal rates of attempted collisions per nucleon for all the employed approaches, not shown), the number of effective collisions per nucleon differs in the two models due to the different treatment of the Pauli blocking, which is more severe in BLOB, owing to the nucleon wave packet extension (for instance, at $\rho^0=0.16$~fm$^3$, large Pauli rejection rates, equal to 98\% in BLOB and to 95\% in SMF, result in different effective collision rates, see discussion in appendix~\ref{sec_appendix_metrics}).
	The main difference emerging from the comparison in Fig.~\ref{fig_collisionvariance} is that the variance of the number of effective collisions per nucleon 
%keeps a dependence on density reduced by a factor $\Ntest$ with respect 
amounts to the mean value reduced by a factor $\Ntest$ in the SMF case, while it equals the mean without any dependence on $\Ntest$ in the BLOB case (several values of $\Ntest$ produce equal results).
	Such study confirms that, while in SMF fluctuations are strongly reduced in proportion to the quantity $\Ntest$, in BLOB fluctuations have large amplitude and exactly equal the mean value, according to the Poisson statistics~\cite{Burgio1991}.
	The quantification of such amplitude is the subject of the following sections.

%%%%%     %%  %%%  %%%     %%%     %%      %%      %%     %%%%%%%%%
%%%%%  %%%%%  %%%  %%  %%%  %%  %%  %%%  %%%%  %%%%%%  %%  %%%%%%%%
%%%%%  %%%%%       %%  %%%  %%  %%  %%%  %%%%    %%%%  %%  %%%%%%%%
%%%%%  %%%%%  %%%  %%       %%     %%%%  %%%%  %%%%%%     %%%%%%%%%
%%%%%     %%  %%%  %%  %%%  %%  %%%%%%%  %%%%      %%  %%  %%%%%%%%

\section{Strategy: comparing BL approaches in nuclear matter to the Fermi-liquid behaviour  \label{sec_BLNM}}
%\section{The Boltzmann-Langevin equation applied to nuclear matter \label{sec_BLNM}}
%
	The purpose of stochastic one-body approaches with collisional correlations like SMF or BLOB is introducing aspects of the Fermi liquid behaviour, including
fluctuations~\cite{Lifshitz1958,Pines1966}, in the description of heavy-ion collisions~\cite{Pethick1988}.

	In the following, we check how Eq.~(\ref{eq:BLOB}) handles isoscalar and isovector fluctuations of the one-body density, in equilibrated nuclear matter, with the aim of demonstrating that its implementation is better suited than approximate methods, like SMF,  to sample the development of inhomogeneities (equivalent process to fragment formation in finite open systems) and the related observables.
	We therefore compare results obtained with BLOB and SMF.

\subsection{Fluctuations in nuclear matter: \\analytic estimate}

%	The purpose of stochastic one-body approaches with collisional correlations like SMF or BLOB is sampling aspects of the behaviour of Fermi liquids~\cite{Lifshitz1958,Pines1966} and including them in the description of heavy-ion collisions~\cite{Pethick1988}.
%	In the following we focus on calculations in nuclear matter.
%
%	In particular, we define the system in a periodic cubic box of edge size $L=39$~fm, and we subdivide it in a lattice of cubic cells of edge size $l$ where we calculate density variances.
%	We initially define the system imposing a uniform-matter effective field $U^0(\vecp)$ which does not depend on configuration space; correspondingly we define a uniform-matter density $\rho^0(\vecp)$ and distribution function $f^0(\vecp)$ depending only on momentum space.
%	We initially define the system imposing a uniform-matter effective field $U^0(p)$ depending only on momentum magnitude, and a corresponding effective hamiltonian $\epsilon(p) = h^0(p) = p^2/(2m) + U^0(p)$.
%	Accordingly, the phase-space distribution function $f^0(\vecp)= \{1+\exp[\epsilon(p)-\mu]/T\}^{-1}$, also not depending on configuration space, is the Fermi-Dirac equilibrium distribution at temperature $T$ for a chemical potential $\mu$.
%

	Let us consider nuclear matter at low temperature.
	Either from the stochastic fluctuating residual term of the BLOB treatment or from an external stochastic force in the SMF approach, we introduce a small disturbance in uniform matter 
$\delta f(\vecr,\vecp,t) = f(\vecr,\vecp,t) - f^0(\vecp,t)$
which lets fluctuations develop in time around the mean trajectory $f^0$.

	By considering neutron and proton distribution functions, we can further decompose fluctuations in isoscalar modes
$\delta f^\s$ and isovector modes $\delta f^\v$
% $\delta f^\s = (f_\n-f_\n^0)+(f_\p-f_\p^0)$ and isovector modes $\delta f^\v = (f_\n-f_\n^0)-(f_\p-f_\p^0)$, 
\begin{eqnarray}
\delta f^\s &=& (f_\n-f_\n^0)+(f_\p-f_\p^0)\\
\delta f^\v &=& (f_\n-f_\n^0)-(f_\p-f_\p^0)\;,
\end{eqnarray}
corresponding to phase-space density modes where neutrons and protons oscillate in phase or out of phase, respectively.
	The temporal evolution of both those modes is obtained by applying the BL equation (\ref{eq:SMF}) to the phase-space fluctuations.
	For symmetric matter, and retaining only first-order terms in 
$\delta f^\q$, one obtains:
\begin{equation}
	\frac{\partial\delta f^\q}{\partial t} + \frac{\vecp}{m}\cdot\nabla_\vecr\delta f^\q - \frac{\partial f^0}{\partial \epsilon}\frac{\partial\delta U^\q}{\partial \rho^\q}\frac{\vecp}{m}\cdot\nabla_\vecr\delta\rho^\q =   \frac{\partial f^0}{\partial \epsilon}\frac{\vecp}{m}\cdot\nabla_\vecr U'
	\;,
\label{eq:BLE_fluctuations}
\end{equation}
where the index $\q$ stands either for isoscalar ($\q=\s$) or isovector ($\q=\v$) modes, $f^0=f_\n^0+f_\p^0$, $U^q$ is the mean-field potential in the $q$ channel  and
$U'$ is an external stochastic force (SMF) or a fluctuating stochastic field (BLOB).
	We dropped the average collision term $I_{\textrm{UU}}$ because we consider small temperatures. %and higher order terms involving density gradients.

	To build our stochastic descriptions we assumed that, at least locally, fluctuations have a small amplitude around their mean trajectory so that $\delta f^\q\!\ll\!f^\q$.  % $\delta f^\q \ll f^\q - \delta f^\q$.
%%	Under this assumption, the model can be analysed in the framework of the linear response theory.
%
	When the system is described as a periodic box, %unstable 
collective modes are associated to plane waves of wave number $\veck$.
	In this case, by expanding on plane waves expressed in Fourier components, we can study the evolution in time of phase space density fluctuations
%$\delta f^\q(\vecr,\vecp,t)=\sum_k \exp(i\veck\!\cdot\!\vecr) f_k^\q(\vecp,t) = \sum_k \exp(i\veck\!\cdot\!\vecr) \exp(i\omega_k\vecr) f_k^\q(\vecp)$ 
%$\delta f^\q(\vecr,\vecp,t)=\sum_k e^{(i\veck\!\cdot\!\vecr)} f_k^\q(\vecp,t) = \sum_k e^{(i\veck\!\cdot\!\vecr)} e^{(i\omega_k t)} f_k^\q(\vecp)$ 
\begin{equation}
	\delta f^\q(\vecr,\vecp,t)=\sum_k e^{(i\veck\!\cdot\!\vecr)} f_k^\q(\vecp,t) 
	= \sum_k e^{(i\veck\!\cdot\!\vecr)} e^{(i\omega_k t)} f_k^\q(\vecp)
\end{equation}
and undulations in the density landscape 
%$\delta\rho^\q(\vecr,t)=\sum_k \exp(i\veck\!\cdot\!\vecr) \rho_k^\q(\vecp,t)$.
$\delta\rho^\q(\vecr,t)=\sum_k e^{(i\veck\!\cdot\!\vecr)} \rho_k^\q(t)$.
	Rewritten in Fourier components, Eq.~(\ref{eq:BLE_fluctuations}) takes the form
\begin{equation}
%	\frac{\partial f_k^\q}{\partial t} 
	i\omega_k f_k^\q
+ i \veck\cdot\frac{\vecp}{m}f_k^\q - i \frac{\partial f^0}{\partial \epsilon}\frac{\partial U_k^\q}{\partial \rho^\q} \veck\cdot\frac{\vecp}{m}\rho_k^\q = i \frac{\partial f^0}{\partial \epsilon}\veck\cdot\frac{\vecp}{m}\F_k^\q
	\;,
\label{eq:BLE_Fourier}
\end{equation}
where $U_k^\q$ and $\F_k^\q$ are Fourier components of the potential $U$ and of the stochastic fluctuating field $U'$, respectively.

% --- Stable matter

%	When only stable modes can propagate, 
{\it When the fluctuation modes are of stable nature,}
the response of the system to the action of the stochastic 
fluctuating field $\F_k^\q$ determines the equilibrium variance $(\sigma_k^\q)^2$ associated to the fluctuation $\rho_\veck^\q$.
	The inverse Fourier transform of $(\sigma_k^\q)^2$ 
gives the equilibrium variance of spacial density correlations 
%$(\sigma_{\rho^\q})^2 \equiv \langle (\delta\rho^\q(\vecr))^2 \rangle = (2\pi)^{-3}\sum_\veck(\sigma_k^\q)^2 d\veck$
\begin{equation}
	(\sigma_{\rho^\q})^2 \equiv \langle (\delta\rho^\q(\vecr))^2 \rangle 
	= (2\pi)^{-3}\sum_\veck(\sigma_k^\q)^2 d\veck
\end{equation}
in a cell of volume $\Delta V$ at temperature $T$. %and for stable modes.
%{\bf [(10) questa definizione e' generale, non solo per i modi stabili}
	At equilibrium, when the level density $\N\equiv(g/h^3)\int \partial_\epsilon f^0 \diff\vecp$ for a degeneracy $g$ can be defined, these variances are related to the curvature of the free energy density %fluctuating field 
$F^\q(k)$ 
through the fluctuation--dissipation theorem so that
\begin{equation}
	(\sigma_k^\q)^2 = \frac{T}{F^\q(k)}\;; \;\;\; (\sigma_{\rho^\q})^2 = \frac{T}{\Delta V}\Big\langle\frac{1}{F^\q(k)}\Big\rangle_\veck
	\;,
\label{eq:fluctuation_dissipation}
\end{equation}
%where $F^\q(k)=\frac{\partial U_k^\q}{\partial \rho^\q}+\frac{1}{\N}$.
where $F^\q(k)= \partial_{\rho^\q} U_k^\q + 1/\N$, and for an average $\langle\cdot \rangle_\veck$ extending over all $\veck$ modes.

On the other hand, for unstable modes, 
the diffusion coefficient $\mathcal{D}$, or rather its %$D_k$ 
%related to a phase-space cell $\Delta V_f$ and projected
projection
on a given unstable mode $k$,  $D_k$,  
determines the following evolution for the intensity of response $(\sigma^\q_k)^2$ for the wave number $k$~\cite{Colonna1993,Colonna1994_a}:
\begin{equation}
	(\sigma^\q_k)^2(t) \approx  D_k\tau_k(e^{2t/\tau_k}-1)+(\sigma^\q_k)^2(t\!=\!0)e^{2t/\tau_k},
\label{eq:diffusion}
\end{equation}
where both the initial fluctuation seeds $(\sigma^\q_k)^2(t\!=\!0)$ and the fluctuation continuously introduced by the collisional correlations contribute to an exponential amplification of the disturbance, characterised by the growth time $\tau_k$.

\subsection{Scenarios for isovector and isoscalar fluctuations}

	In the following, starting from Eq.~(\ref{eq:BLE_Fourier}), 
we concentrate on the propagation of isovector modes, which are always of stable nature, and
isoscalar modes, with a special focus on unstable conditions.

%we select two very instructive situations which are isovector modes in uniform matter and, successively, isoscalar fluctuations in unstable matter.
%	Translated into a violent nuclear-collision scenario, 
%	The first situation, 
	Isovector fluctuations, based on Eq.~(\ref{eq:fluctuation_dissipation}) and studied in Sec.~\ref{sec_ivfluctuations} for nuclear matter at several density values, define how isospin distributes among different phases and portions of the system. 
On the other hand, isoscalar fluctuations developing in mechanically unstable nuclear matter, which rely
 on Eq.~(\ref{eq:diffusion}), studied in Sec.~\ref{sec_isfluctuations}, coincide with the process of separation of those portions of the system into fragments.
	The latter scenario has been intensively investigated~\cite{Chomaz2004} foremost because in open dissipative systems, like heavy-ion collisions, it corresponds to a catastrophic process which can lead to the formation of nuclear fragments~\cite{Tabacaru2003,Borderie2008}.
	The size-distribution of fragments and their formation time are
ruled by the dispersion relation for wavelengths related to unstable $k$ modes so that, when unstable modes succeed to get amplified, inhomogeneities develop and eventually lead to mottling patterns at later times.
	Then, in this case, isovector fluctuations define the isotopic features of fragments 
and their connection to the symmetry energy~\cite{Colonna2013}.
\section{Results on isovector fluctuations
\label{sec_ivfluctuations}}

% Calculations in a cubic periodic box of side $L=$39fm with BOXSMF and BLOBLOB.
% 40test-particles per nucleon are employed.
% Stable symmetric matter is simulated by keeping only the isovector contribution in the potential.
% A free cross section with a threshold at 50mb is used like in SMF, potential-boundary treatments are switched off.
% The temperature is 3MeV, but it then stabilises at larger values.

	Isovector effects in nuclear processes may arise from different mechanisms~\cite{Baran2012,DiToro2003}, like the interplay of isospin and density gradients in the reaction dynamics, %isospin transport, 
or nuclear cluster formation, or the decay scheme of a compound nucleus.
%	Alternately, in some circumstances, a role is played also by isospin distillation~\cite{Chomaz2004,Baran2005}, a mechanism which consists in concentrating the nucleon fraction in the most volatile phase of the system as an effect of the potential term in the symmetry energy~\cite{Colonna2008}.
	In systems undergoing a nuclear liquid-gas phase transition, a role is played also by isospin distillation~\cite{Chomaz2004,Baran2005}, a mechanism which consists in producing a less symmetric nucleon fraction in the more volatile phase of the system along the direction of phase separation in a $\rho_n$--$\rho_p$ space, as an effect of the potential term in the symmetry energy~\cite{Ducoin2007,Colonna2008}.

	Along with these analyses, it is particularly instructive to investigate the developing of isovector fluctuations, around the mean trajectory, in two-component nuclear matter.
%	Those latter correspond to phase-space density modes where neutrons and protons oscillate out of phase.
%	In processes where fragments arise rapidly, like in first-order phase transitions, isovector fluctuations contribute in determining the isospin properties, such as isotopic distributions,  of the low- and high-density fractions which compose the mixed phase.

\subsection{Preparation of a stable and uniform system}

	We consider nuclear matter with periodic boundary conditions. 
We refer the reader to the appendix~\ref{sec_appendix_parameters} for details on the parameters chosen for the calculations in the following sections.

	Selecting isovector modes ($\q\rightarrow \v$) in Eq.~(\ref{eq:BLE_Fourier}), the phase-space density corresponds to $f^\v=f_\n-f_\p$.
%	In order to isolate the isovector behaviour, we concentrate on nuclear matter in stable conditions.%
%	Thus, to keep nuclear matter uniform at any density value (no inhomogeneities will arise in configuration space), 
	In order to select the isovector behaviour, 
we keep only the isovector contribution in the nuclear potential. 
Indeed, in absence of isoscalar terms, the system is stable at all density values $\rho_0$
and one can investigate how isovector fluctuations depend on $\rho_0$. 
% $U^\q\rightarrow U^\v = 2[(\rho_n-\rho_p)/\rho^0]E_{\textrm{sym}}^{\textrm{pot}}$, 
\begin{equation}
	U^\q\rightarrow U^\v = 2[(\rho_n-\rho_p)/\rho^0]E_{\textrm{sym}}^{\textrm{pot}}\;, 
\end{equation}
where $\rho^0$ is the uniform-matter density and $E_{\textrm{sym}}^{\textrm{pot}}$ is the potential term in the symmetry energy.
	Following the procedure of ref.~\cite{Colonna2013}, $U_k^\v$ is obtained from the above quantity by introducing an interaction range through a Gaussian smearing $g_\sigma$ of width $\sigma$, and by taking the Fourier transform; its derivative with respect to $(\rho_n\!-\!\rho_p)$ yields 
%$F^\v(k) = (2/\rho^0)E_{\textrm{sym}}^{\textrm{pot}}(\rho^0)g_\sigma(k) + 1/\N$.
\begin{equation}
	F^\v(k) = (2/\rho^0)E_{\textrm{sym}}^{\textrm{pot}}(\rho^0)g_\sigma(k) + 1/\N\;.
\end{equation}
Thus, from Eq.~(\ref{eq:fluctuation_dissipation}), we obtain the relation: 
%between the isovector variance and the  {\it symmetry free energy}
\begin{equation}
F_{\textrm{eff}}^\v = 
	\frac{T}{2\Delta V} \frac{\rho^0}{(\sigma_{\rho^\v})^2} = \frac{T}{2\Delta V} \frac{\rho^0}{\langle [\delta\rho_n(\vecr)-\delta\rho_p(\vecr)]^2 \rangle} 
%= F_{\textrm{eff}}^\v 
	\;,
\label{eq:isovector_variance}
\end{equation}
where $F_{\textrm{eff}}^\v$ can be assimilated to an effective symmetry free energy which, at zero temperature and neglecting surface effects, coincides with the symmetry energy  $E_{\textrm{sym}}(\rho^0)$.
%ideally, should correspond to the potential term in the symmetry energy $F_{\textrm{eff}}^\v \approx C_{\textrm{sym}}^{\textrm{pot}}$.

	In conventional BUU calculations, however, the smearing effect of the test particles introduces a corresponding scaling factor~\cite{Colonna1994_a}, so that 
$F_{\textrm{eff}}^\v \approx \Ntest E_{\textrm{sym}}$.
	Such scaling actually reduces drastically the isovector fluctuation variance produced by the UU collision term.
	In the following, we investigate how the collision term used in the BLOB approach differs from the UU treatment.
	Since the former is not an average contribution and it acts independently of the number of test particles, we expect a larger isovector fluctuation variance.

	To prepare a transport calculation, the system is initialised with a Fermi-Dirac distribution at a temperature $T=3$~MeV; it is then sampled for several values of $\rho^0$ and the potential, restricted to the only isovector contribution, is tested for a stiff and a soft density dependence of the symmetry energy for symmetric matter (see appendix~\ref{sec_appendix_parameters}).

\subsection{isovector fluctuation variance and symmetry energy}
%
%
%	--- FIGURE Fivrho
%
\begin{figure}[b!]\begin{center}
	\includegraphics[angle=0, width=1\columnwidth]{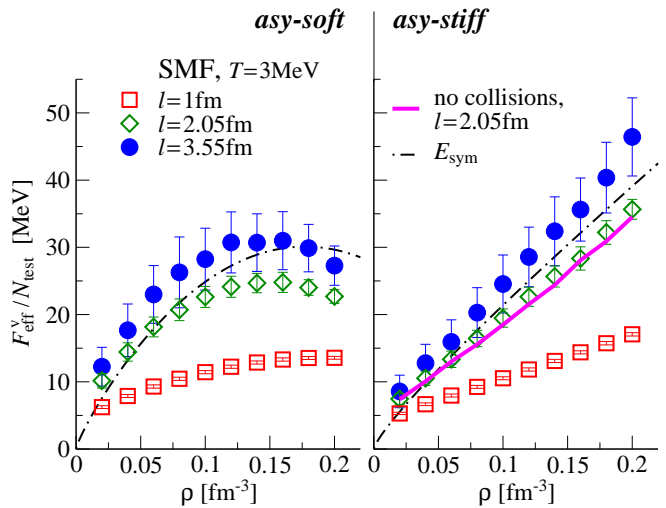}
\end{center}\caption
{
%	Numerical solution of Eq.~(\ref{eq:isovector_variance}) for SMF, including the normalisation by $\Ntest$ for asy-stiff and asy-soft forms of the symmetry energy and for different sizes of the cells where the isovector variance is evaluated.
%	The temperature is extracted for each density bin.
%	For the asy-stiff case a calculation relying only on a collisionless dynamics is added.
%	Error bars indicate the standard deviation of the  $F_{\textrm{eff}}^\v/\Ntest$  distribution obtained for each density bin. 
%	
The effective symmetry energy,  scaled by $\Ntest$, 
as extracted from SMF simulations, using  Eq.~(\ref{eq:isovector_variance}), for nuclear matter at temperature $T$ = 3 MeV and at several density values.
The different symbols correspond to the three edge sizes $l$ of the cells over which the isovector variance is evaluated.
Two parametrizations of the symmetry energy are considered in the calculations:
asy-soft (left panel) and asy-stiff (right panel).
In both panels, the dot-dashed line represents the analytical expression of the symmetry energy.   
%	Comparison between the effective symmetry free energy, scaled by $\Ntest$, calculated with SMF, and the symmetry energy for asy-stiff and asy-soft forms of the symmetry energy (not scaled).
%	The calculation is a numerical solution of Eq.~(\ref{eq:isovector_variance}) for three edge sizes $l$ of the cell over which the isovector variance is evaluated.
	For the asy-stiff case (right panel), an SMF calculation where the collision term is suppressed is also shown for the edge size  $l$ = 2.05 fm
(full violet line). 
%	Error bars indicate the standard deviation of the $F_{\textrm{eff}}^\v/\Ntest$ distribution obtained for each density bin. 
Error bars indicate the standard deviation of the $F_{\textrm{eff}}^\v/\Ntest$ distribution obtained considering several time instants within a large interval at equilibrium.
}
\label{fig_Fivrho}
\end{figure}
%
%
%	--- FIGURE densityvariance versus t
%
\begin{figure}[b!]\begin{center}
	\includegraphics[angle=0, width=.9\columnwidth]{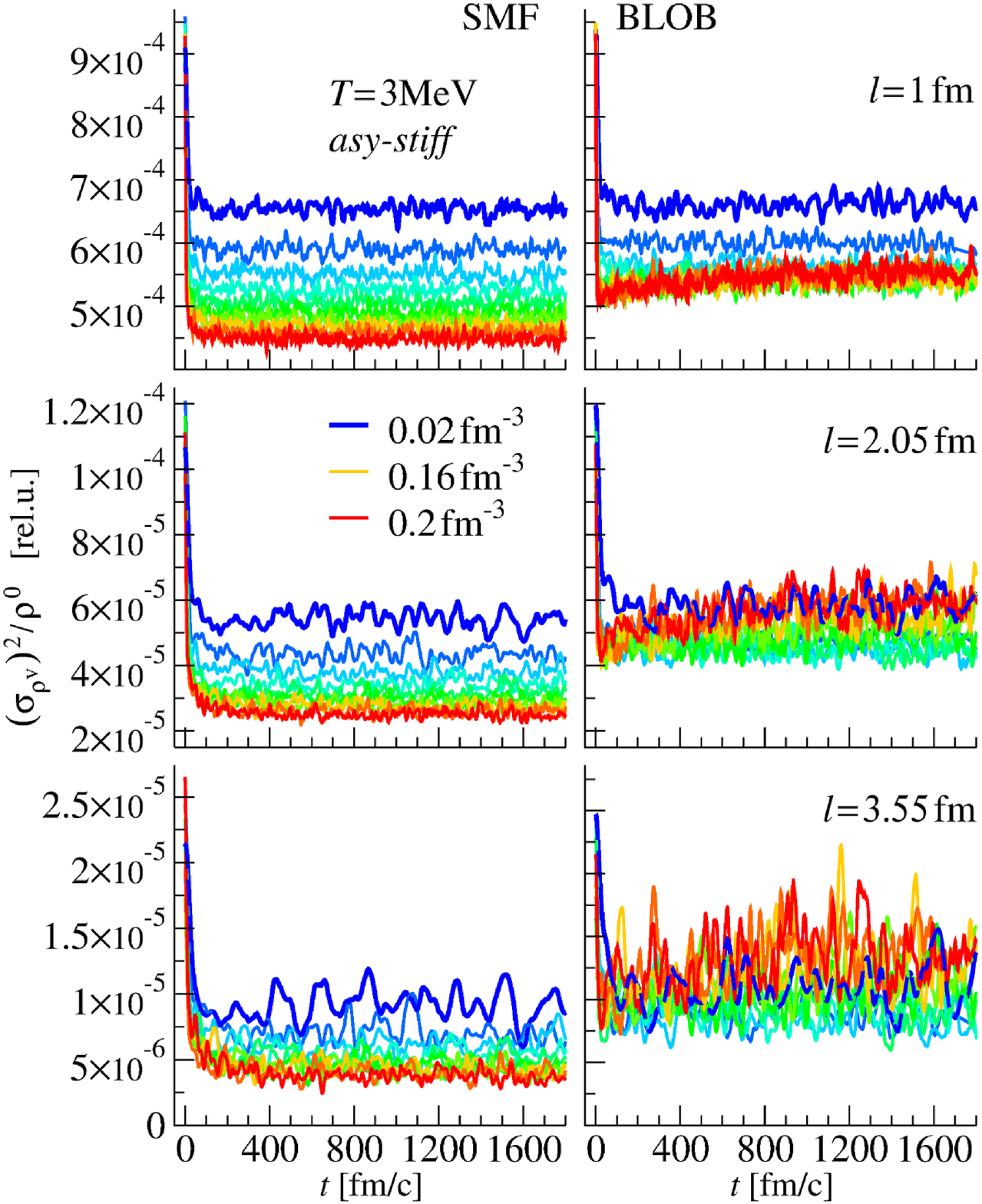}
\end{center}\caption
{
	Isovector variance as a function of time calculated, in a single event, 
with SMF (left) and BLOB (right), 
for different values of the system density, corresponding to the different
colors and ranging from 0.02 fm$^{-3}$ (blue line) to 0.2 fm$^{-3}$ (red line).
Cells of different size $l$ are considered in the calculations: 
$l$ = 1 fm (top panels), $l$ = 2.05 fm (middle panels) and 
$l$ = 3.55 fm (bottom panels). 
An asy-stiff form of the symmetry energy is considered.  
%	BLOB systematically shows larger values and a longer convergence time for large densities.
%	Both calculations converge to the same variance at the small-density limit.
}
\label{fig_densityvariance_versus_t}
\end{figure}

	From a set of calculations for different densities ranging from $\rho^0=0.02$ to $\rho^0=0.2$~fm$^{-3}$ we obtain a numerical solution of the r.h.s. of 
Eq.~(\ref{eq:isovector_variance}) for SMF. We use $\Ntest$ = 40.
	We consider
%the average equilibrium temperature $T_{\textrm{eq}}\approx4.3$~MeV, calculated at saturation density, for all other densities or, alternatively \PN{keep or remove?} 
an equilibrium temperature extracted for each density bin from the slope of the Fermi-Dirac distribution evolved in time.
	The isovector variance $(\sigma_{\rho^\v})^2$, calculated in cells of edge size $l=1$, $l=2.05$ and $l=3.55$~fm, is multiplied by $\Ntest$, in order to extract $F_{\textrm{eff}}^\v$ and to compare it with %the potential term in 
the symmetry energy $E_{\textrm{sym}}$.
	The comparison, shown in Fig.~\ref{fig_Fivrho}, is satisfactory and it is the closest in shape to $E_{\textrm{sym}}$ for larger cells than $l=1$~fm but, however, the large scaling factor $\Ntest$ has to be taken into account.
	The better agreement in larger cells reflects the decreasing importance of surface effects, so that the calculation gets close to the (volume) symmetry energy.

	We notice that an equivalent calculation where the collision term is suppressed yields identical distributions; such collisionless calculation corresponds to switching off the collision term. 
%either in SMF or, equivalently, in BLOB, since the mean-field is implemented identically.
	The need of scaling by $\Ntest$ to recover the expected fluctuation value reflects the fact that isovector fluctuations
are not correctly implemented in SMF.
%thus one just observes the fluctuations related to the use of a finite number of test particles.  
Indeed, much attention is paid in the model to a good reproduction of isoscalar fluctuations and amplification
of mean-field unstable modes, by introducing an appropriate external field \cite{Colonna1998}. 
On the other hand, explicit fluctuation terms are not injected in the isovector channel in SMF.  
	In this case, one just obtains the fluctuations related to the use
of a finite number of test particles which, as far as the Fermi statistics 
is preserved, amount to the physical ones divided by  $\Ntest$.
	These results completes the study of ref.\cite{Colonna2013} concerning SMF.

%	We may remind that, since the mean-field is stochastic, the suppression of the collision term is not compatible with a Vlasov calculation.
%	This analysis indicates that, in this framework, the isovector variance does not come from the collision term.

%
%	--- FIGURE densityvariance versus rho 
%
\begin{figure}[t!]\begin{center}
	\includegraphics[angle=0, width=1\columnwidth]{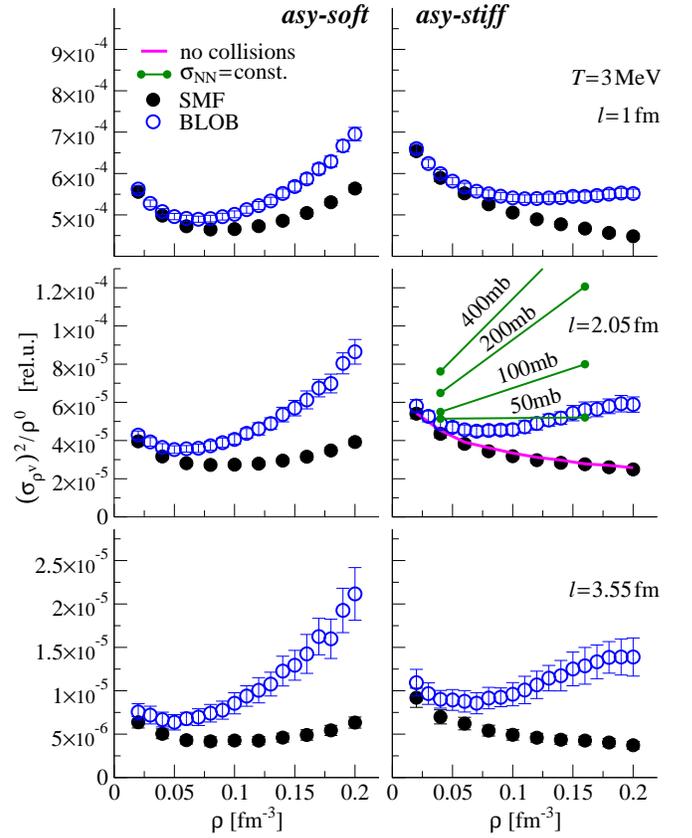}
\end{center}\caption
{
The isovector variance, at equilibrium,  as a function of the system density, 
calculated with SMF (full circles) and BLOB (open circles), for two
parametrizations of the symmetry energy: asy-soft (left panels) and
asy-stiff (right panels) 
Cells of different size $l$ are considered in the calculations: 
$l$ = 1 fm (top panels), $l$ = 2.05 fm (middle panels) and 
$l$ = 3.55 fm (bottom panels). 
  %	Convergence value of isovector variance as a function of system density calculated with SMF and BLOB for asy-stiff and asy-soft forms of the symmetry energy, in cells of different size $l$.
	For one case (asystiff, $l=2.05$~fm) other calculations are shown: SMF with the collision term suppressed (full violet line), and BLOB calculations for different constant values of $\sigma_{\textrm{NN}}$ (green lines).
%	Error bars indicate the standard deviation of the isovector variance distribution obtained for each density bin. 
	Error bars indicate the standard deviation of the isovector variance distribution obtained considering several time instants within a large interval at equilibrium.
}
\label{fig_densityvariance_versus_rho}
\end{figure}

	We now turn to BLOB calculations.  
	Fig.~\ref{fig_densityvariance_versus_t} shows that the isovector variance in BLOB results larger than in SMF. 
%{\it vorrei sopprimere la seguente frase perche' non lo si mostra in fig.5 }
%...and, in general, larger than in a corresponding collisionless calculation.
	Such difference is therefore the effect of the treatment of collisional correlations in BLOB, which displays a dependence with the system density.
	In particular, the low-density limit of the spectrum corresponds to a situation where the collision rate is vanishing.
%producing a Maxwell-Boltzmann statistics: in this case all approaches converge to the same isovector variance.
%{\bf [(12) cosa c'entra la  Maxwell-Boltzmann statistics ?}
	In this case, %when the collision rate is very low, 
the BLOB procedure is practically ineffective (see also the discussion 
in Section V.A)
and all approaches converge to the same isovector variance, just related to the finite number of test particles employed.
	At larger density than saturation ($\rho\ge 0.18$~fm$^{-3}$), BLOB displays a longer path to convergence which is due to the difficulty of relocating large portions of phase space in binary collisions without violating Pauli blocking.

	Fig.~\ref{fig_densityvariance_versus_rho} condenses and extends the information of Fig.~\ref{fig_densityvariance_versus_t} by displaying the density evolution of the isovector variance attained at equilibrium as evaluated in cells of different size $l$, for asy-stiff and asy-soft forms of the symmetry energy.
	The SMF data correspond to those analysed in Fig.~\ref{fig_Fivrho}.
	The BLOB spectra progressively deviate from SMF data for increasing density.
	Such deviation % and the absolute value of the variance, 
increases for larger cell sizes indicating that the isovector fluctuations are better built in large volumes~\cite{Rizzo2008}.
	This is related to the variety of configurations, concerning 
shape and extension of the nucleon wave
packet, which occur in the implementation of the fluctuating collision
integral. This introduces a smearing of fluctuations on a scale comparable
to the wave-packet extension in phase space.  
%fluctuations are not well constructed in fixed cells of small size.}
	However, the gain in isovector variance exhibited by the BLOB approach, indicates that the dependence on $\Ntest$ is partially reduced with respect to the SMF scheme.

\subsection{Interference between mean-field propagation, %numerical 
collisional and numerical correlations \label{sec_ivfluctuations_noise}}

	According to Eq.~(\ref{eq:BLOB}), the BLOB approach should introduce and revive  fluctuations continuously.
%	However, the procedure has chances to work only if there are no other antagonist sources which may destroy the correlations built by the
%collision integral. 
	The agglomeration procedure employed in BLOB is actually able to construct agglomerates of test particles of the same isospin species and which are located around local density maxima in random selected phase-space cells: this technique should preserve at least partially the isovector correlations in the system, contrarily to the usual BUU technique which smears them out.
	This advance with respect to BUU is however not sufficient because of the concurrent effects associated with the %stochastic 
mean-field dissipation.  
Indeed, fluctuations are propagated according to a total inverse relaxation time
%$1/\tau_k' = 1/\tau_k^{\textrm{coll}} + 1/\tau_k^{\textrm{m.f.}}$.  
\begin{equation}
	1/\tau_k' = 1/\tau_k^{\textrm{coll}} + 1/\tau_k^{\textrm{m.f.}}\;,  
\end{equation}
	so that, if the collisional rate is too small, they are damped by the mean-field dynamics 
before they can reach a sizeable amplitude.

Moreover, even in absence of any explicit fluctuation seed, the dynamics is actually affected by a numerical noise, due to the use of a finite number of test particles in the numerical resolution of the transport
equations; such spurious contribution imposes the dependence of $(\sigma_{\rho^\q})^2$ on $\Ntest$~\cite{Colonna1993}.
	If this latter effect may be negligible with respect to the large isoscalar fluctuations 
developing in presence of mean-field instabilities (see next Section),
%introduced by the BLOB stochastic collision term and amplified by the mean-field in presence of instabilities, 
it becomes a highly interfering contribution for the isovector modes.
In other words, the numerical noise leads to an effective 
diffusion coefficient $D_k' = D_k + D_k^{\textrm{noise}}$. %and an effective inverse relaxation time
%$1/\tau_k' = 1/\tau_k + 1/\tau_k^{\textrm{noise}}$.  
%$\tau_{\textrm{noise}}$ depends on the test-particle number (becoming infinite when the test-particle number goes to infinity). 
If a small number of test particles is considered,  and two-body collisions
are not so frequent, then $D_k^{\textrm{noise}}$ prevails over $D_k$, causing a deviation of the fluctuation
amplitude from the correct value.

%	As discussed in ref.~\cite{Reinhard1992}, the approximate mapping of the one-body distribution function through the use of a finite number of test particles, induces a numerical noise that may even cause a deviation
%from the fermionic statistics, towards a classical behaviour of the system.  This effect is more pronounced
%when the collision integral is neglected.  
%	Indeed, this latter contains explicit Pauli-blocking factors and helps
%restoring the fermionic behaviour. 
%%The numerical noise leads to fluctuations corre\Ntestsponding
%%to the expected value, but reduced by $\Ntest$ (as far as the Fermi statistics is still preserved).  

%Moreover, the collision rate has to be confronted with the mean-field propagation time.
%For a low collision rate, the mean-field dissipation will prevail and reduce the fluctuation amplitude. 

%%and   $1/\tau_k^{\textrm{noise}}$ over $1/\tau_k$.
%%%% \tau_noise potrebbe anche non dipendere dal numero di particelle test. 
%% e' un tempo di rilassamento veloce, verso la statistica classica.    
For these reasons, though in principle the fluctuation
equilibrium value, as deduced from BLOB, should not depend on the details of the nucleon-nucleon cross section $\sigma_{\textrm{NN}}$ and on the number of test particle employed, 
our results depend significantly on both ingredients.

	Two ways can be tested to get a deeper insight into
this problem: either the collision term should be considerably enhanced, or fluctuations generated by 
%the mean-field 
test particles should be prevented.

	The first solution can be achieved by simply multiplying $\sigma_{\textrm{NN}}$ by a large factor, with the drawback of then handling incorrect collision rates.
	Of course, this is not a problem if one is interested in equilibrated
matter, as in the present case,  
but it would be crucial when dealing with non-equilibrium processes, 
such as nuclear reactions.
	Some tests in the first direction are proposed in Fig.~\ref{fig_densityvariance_versus_rho}, by employing a constant $\sigma_{\textrm{NN}}$ with progressively larger values, showing that the isovector variance grows with the collision rate, as we expect on the basis of the arguments discussed above.
% suggesting that the above considerations are consistent.

%
%	--- FIGURE Ntest 
%
\begin{figure}[b!]\begin{center}
	\includegraphics[angle=0, width=.95\columnwidth]{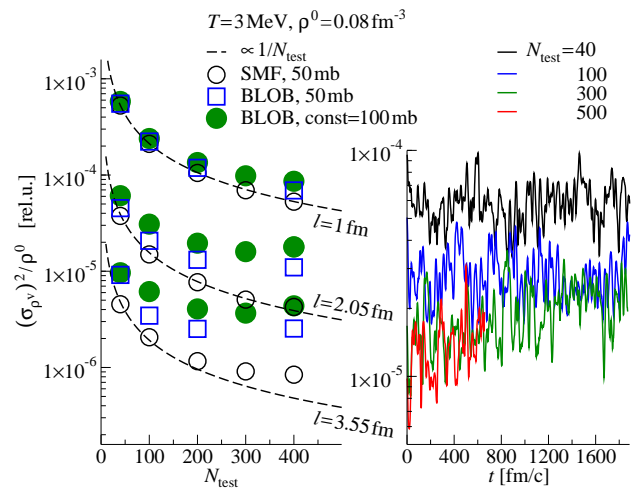}
\end{center}\caption
{
	Left: 
 The isovector variance, at equilibrium, as a function of the number of test
particles, $\Ntest$, as obtained in SMF simulations (open circles) and in BLOB calculations (open squares), using a constant $\sigma_{\textrm{NN}}$=50mb, for nuclear
matter at density $\rho^0$ = 0.08 fm$^{-3}$ and temperature $T=3$~MeV.   
BLOB calculations are also shown for $\sigma_{\textrm{NN}}$=100mb (green full circles).
Cells of different size $l$ are considered, as indicated on the figure. 
Dashed lines represent a fit of SMF simulations, assuming a trend proportional 
to the inverse of the test particle number.   
%$\Ntest$ dependence of the isovector variance at equilibrium calculated with SMF, using constant $\sigma_{\textrm{NN}}$=50mb, and BLOB, using constant $\sigma_{\textrm{NN}}$=50 and 100mb, and compared to a $1/\Ntest$ behaviour.
	Right: 
Time evolution of the isovector variance, as obtained in a single event, for 
%Time dependence showing the tendency to lose $\Ntest$ dependence for 
BLOB calculations with constant $\sigma_{\textrm{NN}}$=100mb and $l=2.05$fm.
The different curves correspond to different number of test particles employed
in the simulations. 
%
%	Study for 
The asy-stiff form of the symmetry energy is considered. 
%, at $\rho^0=0.08$~fm$^{-3}$, in cells of different size $l$.
}
\label{fig_Ntest}
\end{figure}
	The second check would consist in employing the largest possible number of test particles per nucleon.
	In this case, the collisionless transport model would ideally correspond to the Vlasov approach and, when collisional correlations are introduced, interferences with spurious stochastic sources can be highly reduced.
	However, interference effects with the mean-field propagation can still be important, 
depending on the collision rate.
	As far as numerical complexity can be handled, Fig.~\ref{fig_Ntest}, left, illustrates such situation:
SMF calculations show a behaviour $\propto 1/\Ntest$, independently of $\sigma_{\textrm{NN}}$.
	On the other hand, in the BLOB case  
%no $\sigma_{\textrm{NN}}$ dependence, but 
%a pronounced tendency to lose 
%a dependence on $\sigma_{\textrm{NN}}$, as just discussed above.
%but the latter
%tends to diminuish when increasing the number of test particles, whereas the
one observes that,  especially in the
largest cells, where fluctuations are more effective,  
the corresponding variance deviates more and more, for large test particle numbers, 
from the SMF results, reaching a kind of saturation value.  The latter depends on the cell size, $l$,
and on the cross section employed (see also the discussion above). 
%{\it when the number of test particles is increased.}
	Fig.~\ref{fig_Ntest}, right, shows the time evolution of the fluctuation variance. It appears that
for a small number of test particles (up to 100) the variance oscillates around its initial value, which is
essentially associated with the numerical noise and scales as $1/\Ntest$.  Increasing the number of test particles, 
the numerical noise gets smaller and the BLOB fluctuation source prevails on it, building up a fluctuation variance
which is larger than the initial value. 
%illustrates that small, progressively increasing values of $\Ntest$, are related to a systematically decreasing isovector variance, which is still completely dominated by the noise. %mean-field noise.
%	Only when the number of test particles per nucleon becomes very large, the isovector variance loses its dependence on $\Ntest$ and exhibits a clear tendency to grow in time towards a larger value, indicating that isovector correlations are not only preserved, but they are also revived.
	However, since	we are considering systems at low
temperature,
%to apply a linear-response approximation impose to use a not too large temperature, 
the number of nucleon-nucleon collisions is extremely low and insufficient to rapidly introduce a pattern of 
isovector correlations, unless one employes very high values for the cross section: the isovector variance shows in fact a very gentle growth.

	In conclusion, the BLOB fluctuation source term works well in conditions
where the collision rate is large enough, as compared to the mean-field propagation and to the spurious 
dissipative terms associated with the finite number of test particles. 
	These conditions are likely reached in the first, non equilibrated stages of heavy ion collisions, 
%{\it} %at Fermi and intermediate energies, 
but not necessarily for equilibrated nuclear matter at low temperature. In the latter case, the variance associated with the 
fluctuating collision integral can be recovered by artificially increasing the employed $\sigma_{\textrm{NN}}$. 
%(or enhancing the number of test particles). 
Indeed, as shown in Fig.~\ref{fig_densityvariance_versus_rho} and Fig.~\ref{fig_Ntest}, 
we observe that the fluctuation variance built
by BLOB may deviate significantly from the SMF results, being up to a factor ten larger, especially when considering 
fluctuations in larger cells ($l\approx$ 2 to 3 fm). 
\section{Results on isoscalar fluctuations \label{sec_isfluctuations}}

	If fluctuation seeds are introduced in homogeneous neutral nuclear matter at low temperature, Landau zero-sound~\cite{Landau1957} collective modes should stand out and propagate in the system. %~\cite{Pines1966}.
	In the present section we analyse whether the BLOB approach is able to develop, as aimed, isoscalar fluctuations of correct amplitude in nuclear matter spontaneously, and not from an external contribution, when the system is placed in a dynamically unstable region of the equation of state~\cite{Belkacem1994}, like the spinodal zone.
	In this circumstance, as soon as fluctuation seeds are generated, unstable zero-sound waves should be amplified in time.
	In the opposite situation, in conditions of mechanical stability, undamped stable zero-sound waves propagate.
	Then, for stable configurations, the same argumentations of Sec.~\ref{sec_ivfluctuations} hold and, in this case, the fluctuation variance is linked to matter incompressibility.
 
\subsection{Sampling zero-sound propagation in mechanically stable and unstable nuclear matter}

\begin{figure}[t!]\begin{center}
	\includegraphics[angle=0, width=1\columnwidth]{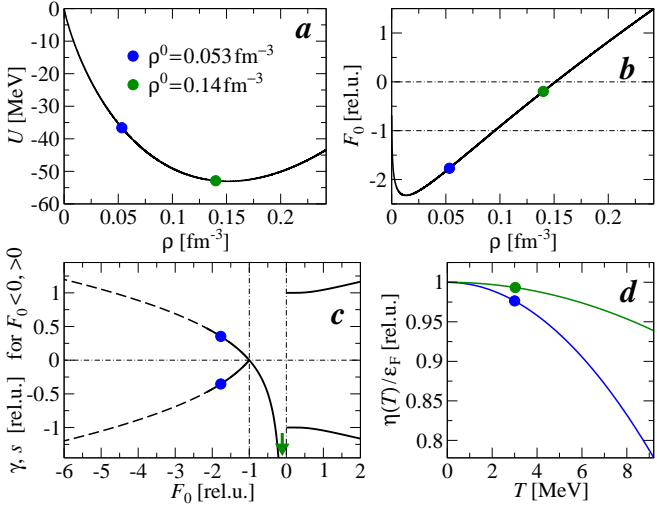}
%	\includegraphics[angle=0, width=.460\columnwidth]{../figs_work/disprelation_analytical/plot_potential.eps}
%	\includegraphics[angle=0, width=.520\columnwidth]{../figs_work/disprelation_analytical/plot_F0.eps}
%\\
%	\includegraphics[angle=0, width=.525\columnwidth]{../figs_work/disprelation_analytical/plot_roots_allowed.eps}
%	\includegraphics[angle=0, width=.455\columnwidth]{../figs_work/disprelation_analytical/plot_F_KT.eps}
\end{center}\caption
{
(a) The isoscalar mean-field potential as a function of the density. The dots indicate
the two density values considered in SMF/BLOB simulations. 
(b) The Landau parameter $F_0$ as a function of the density. 
%	Two choices for the system density are illustrated in relation with corresponding quantities. 
%	(a) Nuclear potential. (b) Landau parameter. 
(c) real (for $F_0>0$) and imaginary (for $F_0<0$) roots of the dispersion relation; the spinodal region corresponds to $F_0<-1$, the region for $-1<F_0<0$ corresponds to Landau damping and positive values of $F_0$ define stable modes.
Blue dots indicate the $\gamma$ values 
corresponding to the density $\rho^0$ = 0.053 fm$^{-3}$.    
The $\gamma$ value corresponding to $\rho^0$ = 0.14 fm$^{-3}$ is beyond the
interval shown in the figure (in the direction of the green arrow).  
%\PN{The arrow indicates that the point for 0.14~fm$^{-3}$ is situated at a very negative value of the ordinate.}
(d) The chemical potential, divided by the Fermi energy, as a function of the temperature $T$, for the two density values considered in the calculations: 
 $\rho^0$ = 0.053 fm$^{-3}$ (blue line) and  $\rho^0$ = 0.14 fm$^{-3}$ (green line).   
The dots correspond to the temperature value ($T=3$~MeV) considered 
in the simulations.   
%Temperature effects on the dispersion relation for the two system densities.
}
\label{fig_disp_rel_analytical}
\end{figure}

	The  propagation %early growth 
of fluctuations in %unstable 
nuclear matter can be described in a linear-response approximation~\cite{Colonna1994_a} as far as deviations from the average dynamical path are small.
	In Eq.~(\ref{eq:BLE_Fourier}), by selecting isoscalar modes ($\q\rightarrow \s$, we drop the $\s$ index in the following), 
% substituting $\partial_t\delta f_k(\vecp,t) = i\omega_k f_k(\vecp,t)$ 
and setting residual contributions to zero, we obtain a linearised Vlasov equation in terms of frequencies $\omega_k$ to describe stable matter with isoscalar contributions:
%	Selecting isoscalar modes ($\q\rightarrow \s$) in Eq.~(\ref{eq:BLE_Fourier}) (we drop the $\s$ index in the following) and setting residual contributions equal to zero, we obtain a linearised Vlasov equation to describe stable matter with isoscalar contributions:
\begin{equation}
	\omega_k f_k + \veck\cdot\frac{\vecp}{m}f_k - \frac{\partial f^0}{\partial \epsilon}\frac{\partial U_k}{\partial \rho} \veck\cdot\frac{\vecp}{m} \rho_k = 0
	\;.
\label{eq:linearised_Vlasov}
\end{equation}
%where we obtained an expression in terms of frequencies $\omega_k$ by substituting $\partial_t\delta f_k(\vecp,t) = i\omega_k f_k(\vecp,t)$.
	Different wave numbers $\veck$ are decoupled, each linked to a collective solution $f_k$ given by the Fourier-transformed equation of motion.
	By applying the self-consistency condition $\rho_k = (g/h^3)\int f_k(\vecp) \diff\vecp$, we obtain the dispersion relation for the propagation of density waves in Fermi liquids at $T=0$ :
\begin{equation}
	1 = \frac{g}{h^3}  \frac{\partial U_k}{\partial\rho} \int \frac{\partial f^0}{\partial\epsilon}  \frac{\veck\cdot\vecp /m}{\omega_k+\veck\cdot\vecp /m} \diff\vecp
	\;.
\label{eq:dispersion_relation_primitive}
\end{equation}
where $\omega_k$ and $-\omega_k$ are pair solutions due to the invariance $\vecp \leftrightarrow -\vecp$.
%\PN{ about eigenfrequency problem...the fact that the solution of the transport eq. only exists for some discrete eigenvalues. i.e. $\omega_k$ is a well defined function of $\veck$...define $U(\veck)$...}
	As well documented in the literature, at $T=0$ eigenmodes $f_k$ depend 
only on states near the Fermi level.
	Indeed, the momentum integral is %should therefore be 
restricted to the Fermi surface because %so that 
$\partial_\epsilon f^0 \approx -\delta(\epsilon-\eF)$, being $\eF$ the Fermi energy.
%and angular and energy dependencies can be decoupled so that % $\vecK\cdot\vecp \approx kp_\textrm{F}$
The dispersion relation reduces to an expression where solutions correspond to sound velocities $s=\omega_k / (k v_\textrm{F})$ in units of Fermi velocity  $v_\textrm{F} = p_\textrm{F}/m$.
	In this case, introducing the Landau parameter 
%$F_0(\veck)=\N_0\partial_\rho U_k= (3/2)(\rho^0/\eF)\partial_\rho U_k$,
\begin{equation}
	F_0(\veck)=\N_0\partial_\rho U_k= (3/2)(\rho^0/\eF)\partial_\rho U_k\;,
\end{equation}
where $\N_0 = \N(T=0)$ is linked to the number of levels at Fermi energy $\eF$, the dispersion relation takes the form~\cite{Kalatnikov1958}:
\begin{equation}
%	\frac{s}{2} \textrm{ln} \left(\frac{s+1}{s-1} \right)  = 1+\frac{1}{F_0}
	1 + \frac{1}{F_0} = L(s) = \frac{s}{2} \textrm{ln} \left(\frac{s+1}{s-1} \right)
	\;,
\label{eq:disprel_T0}
\end{equation}
where the  Lindhard function, $L(s)$ has been introduced.  
%the dependence on $k$ has been removed by the introduction of the sound speed $s v_\textrm{F} = \omega_k/k$, equal for all $k$ waves.
	In correspondence to the effective interaction employed, reflected in the potential shown in Fig.~\ref{fig_disp_rel_analytical}a,
	the Landau parameter is illustrated in Fig.~\ref{fig_disp_rel_analytical}b, while Fig.~\ref{fig_disp_rel_analytical}c presents the roots of the dispersion relation, corresponding to the effective interaction. 

%
%	--- FIGURE dispersion relation analytical
%
\subsection{Warm systems and interaction range}

	Eq.~(\ref{eq:disprel_T0}) is only valid at zero temperature. 
	When the temperature is significant, two-body collision rates become prominent and these mean-field dominated zero-sound waves are absorbed and taken over by hydrodynamical first-sound collective modes.
	Since our approach exploits two-body collisions to introduce fluctuations in a self-consistent mean field, we expect the possible occurrence of a zero-to-first-sound transition which, at variance with other Fermi liquids~\cite{Abel1966}, should be even smeared out due to the small values taken by the Landau parameter $F_0$ in nuclear matter.
	It was found that, depending on how the system is prepared and on the type of collective motion, such transition should arise in a range of temperature from 4 to 5 MeV and occur as late as 200~fm/$c$~\cite{Larionov2000, Kolomietz1996}.
	In practice, zero-sound modes associated to wave vectors $\veck$ characterise the system as long as the corresponding phase velocity exceeds the velocity of a particle on the Fermi surface $\vF$ or, equivalently, as long as the corresponding frequency $\omega_k$ is much higher then the two-body collision frequency $\nu$.
	These premises imply that, after defining a homogeneous initial configuration at a suited finite and not so large temperature, we should study early intervals of time to extract properties of the response function which can be compared with zero-sound conditions.

	Temperature effects can be included, in an approximate manner, considering the low temperature Sommerfeld expansion of the chemical potential $\mu$(T):
%	It was proposed~\cite{Yannouleas1992} that the inclusion of a temperature dependence is still possible in a semiclassical picture. 
%	By taking into account the dissipation of energy from collective to microscopic degrees of freedom~\cite{Blocki1978}, we can introduce the dependence on a finite but small temperature through the ratio between the chemical potential $\mu(T)$ at a temperature $T$ and the Fermi energy $\eF$
%%when supposing that a moving boundary of the system is involved in the dissipation of energy from collective to microscopic degrees of freedom.
%%	Such so-called wall-dissipation model~\cite{Blocki1978} can be applied to a Fermi gas at finite but small temperature $T$ by identifying the moving boundary with the Fermi surface of the system.
%%	This argumentation results in including the ratio between the chemical potential $\mu(T)$ at a temperature $T$ and the Fermi energy $\eF$, which carries the temperature dependence
\begin{equation}
	\frac{\mu(T)}{\eF} \approx 1-\frac{\pi^2}{12} \left(\frac{T}{\eF}\right)^2 
	\;,
\label{eq:wall_dissipation}
\end{equation}
which is  illustrated in Fig.~\ref{fig_disp_rel_analytical}d. %for the two selected densities. 

	As a further modification, we consider that zero-sound conditions also present a strong dependence on the interaction range.
	This latter can be included in the dispersion relation by applying a Gaussian smearing factor of the mean-field potential %$\sigma$, %in units of configuration-space length
which is related to the nuclear interaction range in configuration space~\cite{Colonna1994,Kolomirtz1999}.

\begin{equation}
	U \rightarrow U \otimes g(k), \;\;\textrm{with} \;\; g(k)= \textrm{e}^{-\frac{1}{2} (k\sigma)^2}
	\;,
\label{eq:Gaussian_smearing}
\end{equation}

From Eq.~(\ref{eq:wall_dissipation}) and Eq.~(\ref{eq:Gaussian_smearing}), 
the dispersion relation, Eq.~(\ref{eq:disprel_T0}), involves  
%we obtain a new form of Eq.~(\ref{eq:disprel_T0}) which involves both the dependence on the interaction range and on temperature, throug 
an effective Landau parameter \cite{Colonna1994}, 
\begin{equation}
	\widetilde{F}_0(k,T) = \frac{\mu(T)}{\eF}{F}_0g(k)
	\;.
\label{eq:effectiveF}
\end{equation}
%\begin{equation}
%	\left(\frac{\eta(T)}{\eF}F_0g(k)\right)^{-1} = 1 - \frac{s}{2} \textrm{ln} \left(\frac{s+1}{s-1} \right)
%	\;.
%\label{eq:disprel}
%\end{equation}

	Mechanically unstable conditions are experienced when the evolution of local density $\rho$ and pressure $P$ implies that the incompressibility 
%$\chi^{-1} = \rho \frac{\partial P}{\partial\rho}$ 
%$\chi^{-1} = \rho \partial_{\rho}P$
is negative. 
%in a Fermi liquid at zero temperature ($T=0$) 
	This situation is reflected by an effective Landau parameter $\widetilde{F}_0(k=0,T)$ 
smaller than $-1$, so that   % \cite{Pethick1988}:
\begin{equation}
%	\chi^{-1} = \rho \frac{\partial P}{\partial\rho} = \frac{2}{3}\rho\eF
%	\rho \frac{\partial P}{\partial\rho} = \frac{2}{3}\rho\eF
%[1+\widetilde{F}_0(k=0,T)] \;\; <\,0
	\frac{\partial P}{\partial\rho} \approx \frac{2}{3}\eF
[1+\widetilde{F}_0(k=0,T)] \;\; <\,0
	\;,
\label{eq:chi_Landau}
\end{equation}
and it corresponds to imaginary solutions of the dispersion relation~\cite{Pomeranchuk1959}.
	By replacing $s\rightarrow i\gamma$, the relation yielding imaginary solutions can be put in the form:
% $s\leftarrow i\gamma$ which can be written as (see appendix in \cite{Pethick1988} and recommendations in \cite{Colonna1994})
\begin{equation}
	1 + \frac{1}{\widetilde{F}_0(k,T)}
%\left(\frac{\eta(T)}{\eF}F_0g(k)\right)^{-1} 
= \gamma\,\textrm{arctan}\frac{1}{\gamma}
	\;.
\label{eq:disprel_gamma}
\end{equation}

%where $\gamma = i~s$. 
The growth rate $\Gamma_k=1/\tau_k$ is obtained from the solutions of the dispersion relation 
\begin{equation}
% |\gamma| = |\omega_k|/(k\vF) = \frac{1}{\tau_k k} \frac{m}{\pF}
|\gamma| = \frac{|\omega_k|}{k\vF} = 
%\frac{1}{\tau_k k} \frac{m}{\pF}
\frac{1}{\tau_k k \vF}
	\;,
\label{eq:gamma}
\end{equation}
%\PN{$\hbar$ or not $\hbar$?}

	As far as the Fermi statistics is kept in a sufficiently large periodic portion of mechanically unstable nuclear matter, and a fluctuation source term
is acting,  
%and if the occupancy variance is imposed to be equal to $f(1-f)$ in a phase-space cell, 
the expectation is that the intensity of the response should be amplified with the growth rate 
$\Gamma_k$ 
imposed by the mean-field potential $U$ as a function of the unstable mode $\veck$.

\subsection{Obtaining the dispersion relation \label{sec_disprel}}

	To check such expectation numerically through a BL transport approach 
	we keep the same scheme for the definition of the box metrics as in Sec.~\ref{sec_ivfluctuations}; the isoscalar density variance is calculated over cells of edge size $l=1$~fm.
	We now use the full parametrisation of the energy potential per nucleon Eq.~(\ref{eq:pot}), where we use a stiff density dependence of %$E_{\textrm{pot}}$ 
	the symmetry energy
(the same parametrisation was analysed in ref.~\cite{Colonna1997}).
%	A value of $C_{\textrm{surf}}\!=\!-7/\rho_0$ MeV fm$^5$ is chosen for the surface term.
	Nuclear matter is isospin symmetric and it is initially uniform and prepared at a temperature $T=3$~MeV and at densities equal to $\rho^0\!=\!0.053$ and $0.14$~fm$^{-3}$. %\PN{(define $\rho_{\textrm{sat}}$)}.
	Fig.~\ref{fig_disp_rel_analytical} illustrates the values taken by the potential and by the dispersion relation related to these choices.
%	$\Ntest=40$ test particles per nucleon are employed if not otherwise specified.
	The collision term involves the usual isospin- and energy-dependent free nucleon-nucleon cross section with an upper cutoff at $\sigma_{\textrm{NN}}=50$~mb.

	Within the dynamical calculation
we should register at each interval of time $t$ the density in all cells of edge size $l$ of the lattice which constitutes the periodic system of edge size $L$.
	A specific cell can be identified by the vector $\vecn'$. 
%having components equal to 
%%the number of subdivisions along each configuration-space coordinate.
%the number of nodes of a given undulation along each configuration-space coordinate.
Having introduced such a lattice, the perturbation wave number $k$ 
can be expressed as $k=2\pi n /L$, where $n$ is the modulus of 
a vector ranging from 1 to $n_{\textrm{max}} = L/l$ along each of the three spacial directions.  
%	The magnitude of this vector is related to the mode $k=2\pi n l/L$.
	Then the amplitude of the isoscalar fluctuation of a mode $k$ is obtained from the Fourier transform, $F_\veck$,
of the space density
\begin{eqnarray}
	&&\sigma^2_k(t) = \langle F_\veck^2(t)\rangle = 
	\frac{1}{l^3}\Big\langle\Big[\sum_{\vecn'}\rho_{\vecn'}(t)\,\exp\big(ia\vecn\!\cdot\!\vecn'\big)\Big]^2\Big\rangle \\\notag
	&&\propto \Big\langle\Big[\sum_{\vecn'}\rho_{\vecn'}(t)\textrm{cos}(a\vecn\!\cdot\!\vecn')\Big]^2 \!+\! \Big[\sum_{\vecn'}\rho_{\vecn'}(t)\textrm{sin}(a\vecn\!\cdot\!\vecn')\Big]^2\Big\rangle 
	\;,
\label{eq:isoscalar_amplitude}
\end{eqnarray}
where $a=2\pi  l/L$ and the average is extended over all orientations of $\veck$. 
%{\bf dovrebbe esserci il 2pi * l / L anche nell'argomento di sin e cos}
%which define the position of cells contributing to the mode $k$.

\begin{figure}[b!]\begin{center}
	\includegraphics[angle=0, width=1\columnwidth]{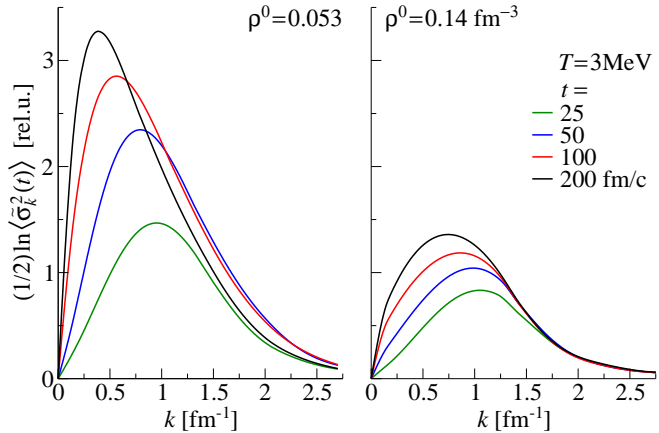}
\end{center}\caption
{
	BLOB calculation. Response intensity $\tilde{\sigma}^2_k(t) = \sigma^2_k(t)/ \sigma^2_k(t=0)$, averaged over 100 dynamical paths,
as a function of the wave number $k$. The different curves correspond to different time instants, as indicated on the figure. 
Left panel: results for $\rho^0\!=\!0.053$~fm$^{-3}$ (spinodal).
Right panel: results for $\rho^0\!=\!0.14$~fm$^{-3}$ (Landau damping). 
}
\label{fig_isoscalar_variance}
\end{figure}

	The distribution of ratios $\tilde{\sigma}^2_k(t) = \sigma^2_k(t)/ \sigma^2_k(t=0)$, averaged over several dynamical paths, is shown at different time intervals in Fig.~\ref{fig_isoscalar_variance} for the two density choices. 
It should be noticed that the initial fluctuation amplitude is due
to the finite number of test particles employed in the calculations. 
However, as soon as the BLOB term starts to act, fluctuations of larger amplitude are built up (see also the discussion in Section V.A), and further amplified by the unstable mean-field. 
	The system prepared at $\rho^0\!=\!0.053$~fm$^{-3}$, inside the spinodal region, exhibits a clear  growth of instabilities as a function of time for some $k$ waves, while the system prepared at $\rho^0\!=\!0.14$~fm$^{-3}$, outside the spinodal region, presents an evolution of the response intensity corresponding to a more gentle growth in a Landau-damping regime.
%which does not lead to the developing of disturbances for any $k$ wave, as we would expect for a phenomenon of Landau damping. 
	However, even in this latter case, fluctuations reach a significant amplitude, owing to the small compressibility value in the density region considered, but cluster formation is not observed.  

%
%	--- FIGURE isoscalar variance
%
%
%
%	--- FIGURE isoscalar variance versus time early
%
\begin{figure}[t!]\begin{center}
	\includegraphics[angle=0, width=.95\columnwidth]{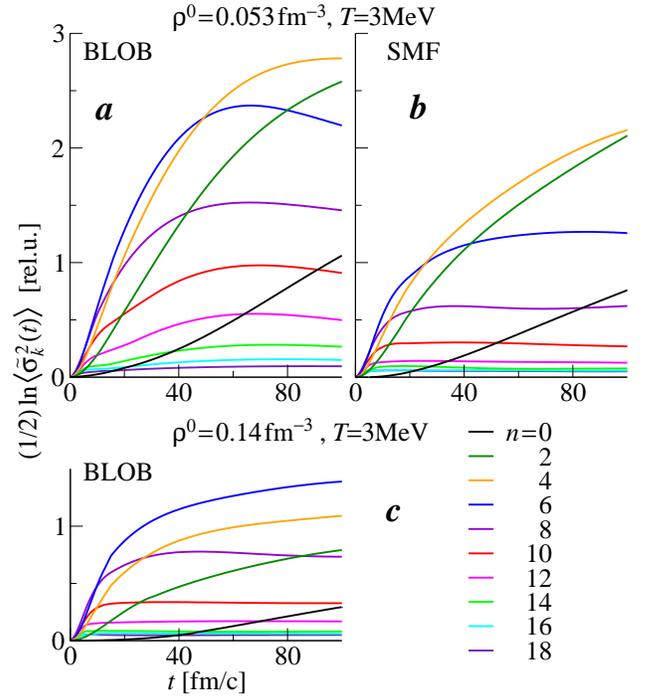}
\end{center}\caption
{
	Early time evolution of the response intensity $\tilde{\sigma}^2_k(t)$ for several modes ($n=j$ stands for all $k$ modes within $2\pi(j-1)/L
\le k< 2\pi j/L$).
	(a) BLOB calculations for $\rho^0\!=\!0.053$~fm$^{-3}$ (spinodal). 
%The leading modes are compared to linear fits.
	(b) SMF calculations for $\rho^0\!=\!0.053$~fm$^{-3}$.
	(c) BLOB calculations for $\rho^0\!=\!0.14$~fm$^{-3}$ (Landau damping).
}
\label{fig_is_variance_early}
\end{figure}
%
%	--- FIGURE isoscalar variance versus time late
%
%\begin{figure}[b!]\begin{center}
%	\includegraphics[angle=0, width=.8\columnwidth]{../figs_work/isoscalar_variance_versus_time_late/f_nwave_BLOB_SMF_stiffsymm_box39_compare.pdf}
%\end{center}\caption
%{
%	Evolution of the response intensity $\tilde{\sigma}^2_k(t)$ over a long interval of time for $n=2$.
%Comparison of SMF and BLOB calculations
%}
%\label{fig_is_variance_late}
%\end{figure}
%
%
%	--- FIGURE growth rate
%
\begin{figure}[t!]\begin{center}
	\includegraphics[angle=0, width=1\columnwidth]{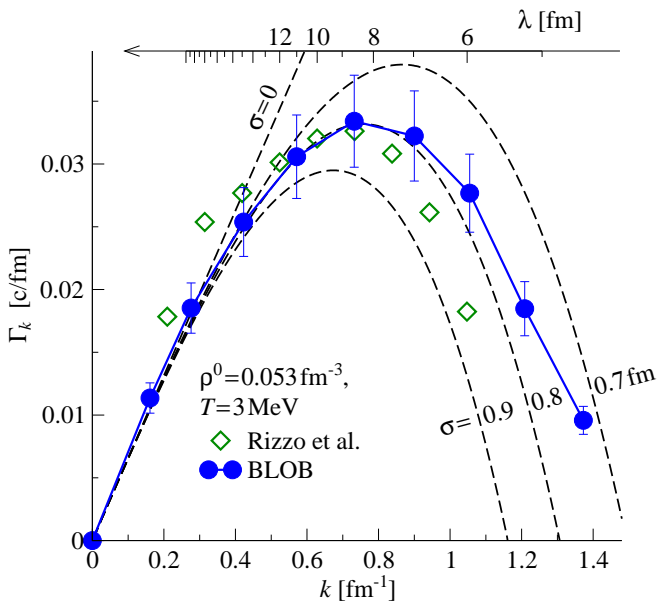}
\end{center}\caption
{
BLOB calculations for nuclear matter at $\rho_0$ = 0.053 fm$^{-3}$ and 
temperature $T$ = 3 MeV: The growth rate $\Gamma_k$ , as extracted from the 
simulations (see Eq.(35)), is plotted as a function of the wave number $k$ (full
line with dots). 
%	Dispersion relation: BLOB calculation compared to the analytic relation of Eq.~(\ref{eq:gamma}).
%	$k$ and $\Gamma$ values are averaged over modes belonging to discrete intervals of $n$.
%	Uncertainties are evaluated from variances around the linear response.   related to the linear fit ???
	Uncertainties are evaluated from the linear-response fit.
The dashed lines correspond to analytical predictions, obtained solving Eq.(32)
with different values of the width $\sigma$ of the Gaussian smearing factor.   
	For comparison, a calculation in one-dimension within the approach of ref.~\cite{Rizzo2008} is also shown (green diamonds).
}
\label{fig_growth_rate}
\end{figure}

	Correspondingly, the early evolution in time of $\tilde{\sigma}^2_k(t)$ is analysed in Fig.~\ref{fig_is_variance_early} for the two density choices. 
%In Fig.~\ref{fig_is_variance_early}a, the leading modes can be compared to linear fits applied to 
%intervals ranging from around 20fm/$c$ to time instants close to saturation %(see Eq.(16)). 
%(see Eq.~\ref{eq:diffusion}).
For the leading modes, one can consider a linear fit of
the quantity plotted in Fig.~\ref{fig_is_variance_early}a, 
%the leading modes can be compared to linear fits applied to 
in intervals ranging from around 20fm/$c$ to time instants close to saturation %(see Eq.(16)). 
(see Eq.~\ref{eq:diffusion}).
	The very initial path is excluded from the fit because, as previously mentioned, the fluctuation mechanism sets in spontaneously after that a sufficient number of collisions has occurred, and does not emerge from suited initial conditions.
	Differences from the ideal linear response in the growing side of single modes indicate a more complex
behaviour, resulting from the coupling of different wavelengths and the tendency toward a chaotic evolution~\cite{Baldo1995}; 
%	\PNout{The comparison is not perfect because the time evolution of single modes may exhibit a more complex behaviour, indicating that different wavelengths get coupled and the dynamics tends to a chaotic behaviour~\cite{Baldo1995};}
%\PNout{such difficulty is taken into account in the large error bars {\bf of Fig.12 ???}.}
	A SMF calculation is also presented for the unstable system, where the linear growth of the leading modes is initially comparable to the BLOB approach and deviates at later times.
%	\PNout{This comparison is completed for a longer interval of time in Fig.~\ref{fig_is_variance_late} for small $k$ modes (for which the Gaussian smearing has no effect on the corresponding dispersion relation): the SMF and BLOB unstable modes grow with slightly different slopes but finally reach the same saturation value for the isoscalar fluctuation variance.}
%
%	This behaviour is due to the different efficiency of the collision term %in preserving the unstable modes in building fluctuations in the two models.  
%	Indeed, the saturation regime %which in absence of perturbations from the mean field would correspond to the same value for the two models,
% is reached earlier in BLOB, because of the more efficient fluctuation source. 
%
	This behaviour is due to the efficiency of the collision term in the BLOB model in reviving fluctuations of correct amplitude, compared to SMF, where fluctuations are not introduced by the collision term.   
	As a consequence, in SMF fluctuations decay by the combination of small wavelengths into larger ones, while in BLOB higher fluctuation amplitudes can be attained before reaching the saturation regime.

	The numerical extraction of the growth rate $\Gamma_k$, 
i.e., of the quantity given by %equivalent to 
the analytic relation of Eq.~(\ref{eq:gamma}) is obtained from the time derivative 
(at early time instants) of the amplitude of the isoscalar fluctuation for a given mode $k$ as
\begin{equation}
\Gamma_k = \frac{1}{2} \frac{\partial}{\partial t} {\textrm ln}\prec \tilde{\sigma}^2_k(t) \succ 
	\;,
\label{eq:tau_numerical}
\end{equation}
where the average $\prec \cdot \succ$ is taken over several stochastic dynamical trajectories.
%{\bf dalla (28) in realta' si estrae 2Gamma}
% the division of the fluctuation amplitude by the initial value has the only purpose of simplifying some calculation procedures.
	Such analysis is presented in Fig.~\ref{fig_growth_rate}, where the numerical calculation, averaged over 100 events, is compared to the analytic result of Eq.~(\ref{eq:gamma}).
	The range of the interaction, as an effect of the implemented surface term, would correspond to a Gaussian smearing of around $\sigma=0.8$~fm to $\sigma=0.9$~fm.
%	The range of the interaction would correspond to a Gaussian smearing of about $\sigma=0.9$~fm, when no surface term is introduced Eq.~(\ref{eq:pot}); the inclusion of the surface term has the effect of reducing the interaction range, yielding a Gaussian smearing of around $0.8$~fm and shifting the ultraviolet cutoff of the $k$-mode spectrum to larger values.
	We infer that BLOB reproduces consistently the expected dispersion relation within the uncertainties of the linear regression.
	Another calculation, also based on the same mean-field, but which employs the earlier approach of ref.~\cite{Rizzo2008}, also solved in three dimensions but with fluctuations developing along one axis of configuration space, produces a similar result.
	While BLOB keeps the different unstable modes decoupled for a more extended interval of time during their early growth, also resulting into a larger ultraviolet cutoff, the other 
approach (green points) presents some alterations due to the combining of unstable modes, where small-wavelength ($n\!>\!7$) are gradually absorbed by large-wavelength ($n\!=\!0,n\!=\!1$).
	The effect in this case is an increase of the growth rate for small $k$ values and it signs the entrance of the chaotic behaviour which characterises larger times~\cite{Jacquot1996}. 
% \PN{Is this comparison with Rizzo interesting? Do we leave it or do we keep it?}
% \PN{add another fig like Fig.~\ref{fig_is_variance_late}, where, over a long time range it is shown how small-wavelength modes ($n>7) are gradually absorbed by large-wavelength modes ($n=0,n=1). }
	The largest $k$ modes, corresponding to wavelengths which drop below the Gaussian smearing width $\sigma$ are meaningless.
	As a final result of this study, the leading modes are found in a wavelength range from 8 to 9fm, and for a growth time $\tau_k$ of around 30fm$/$c. 

% Similar comparisons where done for earlier versions of BLOB leading to similar results~\cite{1WM2014}

%%%%%     %%  %%%  %%%     %%%     %%      %%      %%     %%%%%%%%%
%%%%%  %%%%%  %%%  %%  %%%  %%  %%  %%%  %%%%  %%%%%%  %%  %%%%%%%%
%%%%%  %%%%%       %%  %%%  %%  %%  %%%  %%%%    %%%%  %%  %%%%%%%%
%%%%%  %%%%%  %%%  %%       %%     %%%%  %%%%  %%%%%%     %%%%%%%%%
%%%%%     %%  %%%  %%  %%%  %%  %%%%%%%  %%%%      %%  %%  %%%%%%%%
%\section{Properties of nuclear droplets from isoscalar and isovector fluctuations}
\section{Connecting nuclear matter to open systems \label{sec_open}}
%
%	isovector and isoscalar fluctuation -> their combination
%
	The aim %subject 
of this work is studying the effect of isovector and isoscalar fluctuations. %separately.
	The ultimate purpose of the transport approaches discussed therein is describing the formation of nuclear fragments in a fermionic system and their properties through the combination of these two types of fluctuating modes, as will be detailed more diffusely in forthcoming works.
	In particular, isovector fluctuations, on top of other isospin transport effects, impose that the isospin content is distributed through a density-dependent process of distillation, supplemented by an isotopic variance.
	The onset of isoscalar modes is then responsible for breaking the uniformity of the density landscape and eventually partitioning it into nuclear fragments, where the isospin properties of the initial nesting sites are preserved.
	The isoscalar and isovector mechanisms should therefore be intimately connected in order to describe fragment formation.

%
%	Qualitatively, about isoscalar wavelengths
\subsection{Fragment formation: patterns and time scales}
	Qualitatively, we may underline some connection between the wavelengths involved in the dispersion relation analysed in Fig.~\ref{fig_growth_rate}, and fragment formation~\cite{Matera2000}, considering that at the system density $\rho^0$ the leading modes correspond to fragments of mass $A\approx \rho^0 \lambda^3$; for the leading wavelengths, this corresponds to a distribution of sizes peaked around Neon. 
	These results are also in agreement with other previous studies where quantum effects were taken into considerations explicitly~\cite{Ayik1995,Jacquot1997,Colonna1998bis,Norenberg2000,Ayik2008} despite a more schematic treatment of fluctuations, or of the whole dynamics (2-dimension treatments, fluctuations propagated from an initial state, spherical geometries).
	In this respect, BLOB extends these previous attempts to a model that can be applied at the same time to nuclear matter and, rather successfully, to heavy-ion collisions in three dimensions and without any preliminary initialisation of fluctuation seeds \cite{Napolitani2013,Napolitani2015}.

\begin{figure}[t!]\begin{center}
	\includegraphics[angle=0, width=.85\columnwidth]{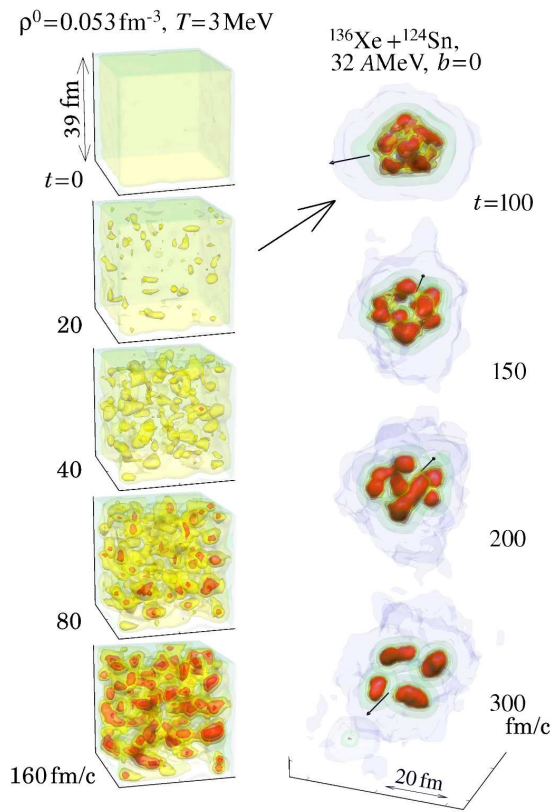}
\end{center}\caption
{
	Density-landscape at several time instants %evolution %in configuration space 
for nuclear matter in a periodic box at $\rho^0\!=\!0.053$~fm$^{-3}$ 
and T = 3 MeV (left) and in a hot open nuclear system formed in a head-on $^{136}$Xe+$^{124}$Sn collision at 32~$A$MeV (right); arrows indicate the beam direction. Both simulations employ the BLOB approach with the same mean-field properties.
%	The arising of spinodal fragmentation characterises both systems in the same interval of time ranging from about $t=80$ to about $160$ fm/$c$. In the heavy-ion system the spinodal signal is shown to get smeared out at later times.
	The big arrow proposes an analogy between nuclear matter and the open system in correspondence with the arising of the spinodal instability.
}
\label{fig_mottling}
\end{figure}

%
%	timing
%
	From the growth time of the leading modes in Fig.~\ref{fig_growth_rate}, we infer that the corresponding process of fragment formation would be rather short, progressing from when the system has been largely diluted.
	This suggests that the scenario studied in nuclear matter can be quite directly translated to the phenomenology of open systems~\cite{Chomaz2004}.
	As an example, Fig.~\ref{fig_mottling} illustrates the correspondence between a portion of nuclear matter (simulated for $T=3$~MeV and $\rho^0\!=\!0.053$~fm$^{-3}$ for an interaction defined as in Eq.~(\ref{eq:pot}) and a hot system formed in the collision $^{136}$Xe$+^{124}$Sn at 32~$A$MeV for a central impact parameter $b=0$ (such system was studied in an experimental campaign~\cite{Moisan2012,Ademard2014}).
%	\PNout{In particular, we remark that fragments arise during comparable intervals of time: the process starts at around 100fm/$c$ by exhibiting a spinodal signal (equal-size fragments), which is then smeared out (fragment recombination) when exceeding about 150fm/$c$.}
	In particular, we observe some analogy between the early time when inhomogeneities emerge in nuclear matter (20fm/$c$) and when fragments start forming in an open system (around 100fm/$c$) right after accessing low-density spinodal conditions (around 80fm/$c$).
	In both systems, a spinodal signal stands out by exhibiting equal-size inhomogeneities in configuration space within a similar time scale~\cite{Napolitani2013}, and it is smeared out by fragment recombination later on.
	At even later times, the evolution is different, in the box calculation clusters continue interacting with each other while in the open system they split apart.
%
%	--- FIGURE  mottling
%
%
%	--- FIGURE distillation
%
%\begin{figure}[b!]\begin{center}
%	\includegraphics[angle=0, width=.8\columnwidth]{../figs_work/distillation/BLOB_XeSn32_Irho.pdf}
%\end{center}\caption{
%	Average isovector density as a function of average local density measured in potential ripples in an early time span (before fragment formation) and in a late time interval (during fragment formation)
%	Sets of potential-ripple sizes are indicated by the corresponding mass $A$.
%}
%\label{fig_distillation}
%\end{figure}

%
%	open systems
%
	Experimental investigations of heavy-ion collisions at Fermi energies already pointed out that the range of masses given by the dispersion relation of Fig.~\ref{fig_growth_rate} is actually favoured in multifragmentation mechanisms; the kinematics of the process was also found to be rather explosive.
	The spinodal mechanism was therefore proposed as a suited description~\cite{Borderie2001,Desesquelles2002}; 
	BLOB has already been adapted successfully to nuclear collisions and tested over various systems which experience spinodal instability~\cite{Napolitani2013,Napolitani2015,Colonna2016}.
%	\PNout{In particular, the timing of the process was found to be rather short and simulations of heavy-ion collisions at Fermi energies employing the BLOB approach described the disassembly of the system as progressing from around 100fm/$c$, when the lowest density conditions are experienced in the system, to about 130fm/$c$ when inhomogeneities stand out in the configuration space; those latter would however separate into fragments at later times.}

\subsection{Isospin content in fragments}

\begin{figure}[b!]
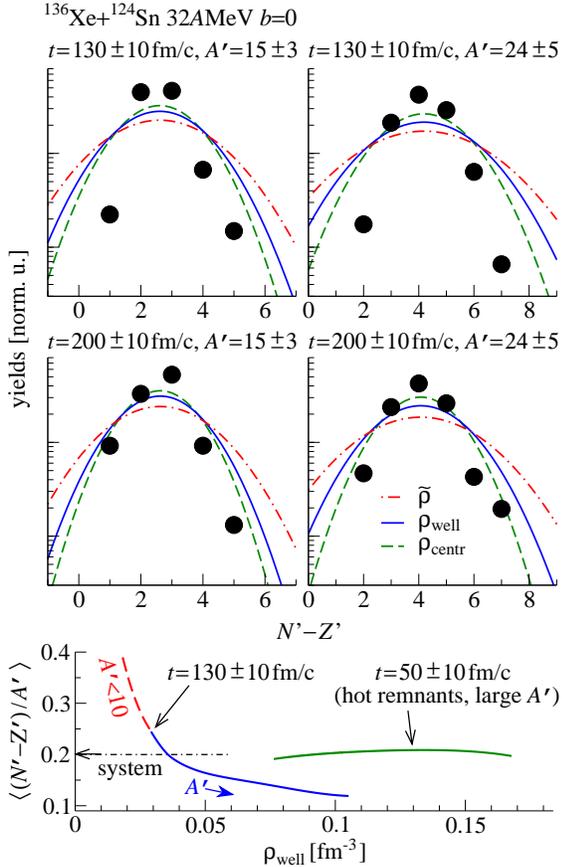
\begin{center}
	\includegraphics[angle=0, width=.85\columnwidth]{fig11a}\\
	\includegraphics[angle=0, width=.85\columnwidth]{fig11b}
\end{center}\caption{
	BLOB simulation of head-on $^{136}$Xe+$^{124}$Sn collisions at 32~$A$MeV.
	Upper row:
Isotopic distribution (full dots) in potential ripples containing $N'$ neutrons and $Z'$ protons for forming-cluster configurations with masses around $A'=15$ (left) 
and $A'=24$ (right)  at t = 130 fm/c.
The lines correspond to analytic distributions (see Eq.(36)), corresponding to 
the density extracted from different portions of potential ripples (see text). 
Middle row: The same as in the upper row, but at t = 200 fm/c. 
	Bottom. Average isospin content measured in potential ripples at two different times, as indicated on the figure,  as a function of $\rho_{\textrm{well}}$. Larger density values correspond to pre-fragments of larger mass.
The dot-dashed line with an arrow indicates the asymmetry of the projectile-target composite system. 
%IMF mass $A'$ evolves according to a process of isospin distillation.
}
\label{fig_isotopicdistributions}
\end{figure}

	Fig.~\ref{fig_isotopicdistributions} completes the survey, investigating the isospin content in potential ripples containing $N'$ neutrons and $Z'$ protons for the system $^{136}$Xe$+^{124}$Sn at 32~$A$MeV.
	Distributions of isotopic variances are calculated for the most probable mass range around a forming Carbon and a forming Neon (two upper rows).
	The distributions are studied in an early time span (before that
fragments are clearly formed, %fragment formation 
around 130fm/$c$) and in a late time interval (during fragment formation around 200fm/$c$). 
	They are compared with the analytic distributions obtained at a temperature $T=5.5$~MeV, as extracted from the calculation, and at the local density $\rho$.
	The isotopic variance (see Eq.~(\ref{eq:fluctuation_dissipation})) can be studied as the probability of variation $\delta$ around the mean value of $N'-Z'$ and for a given $A'$ yielding the distribution 
%$Y \approx \textrm{exp}[-(\delta^2/A')\,C_{\textrm{sym}}(\rho)/T]$
%(see Eq.~(\ref{eq:fluctuation_dissipation}) for the variance).
\begin{equation}
	Y \approx \textrm{exp}[-(\delta^2/A')\,C_{\textrm{sym}}(\rho)/T]\;.
\end{equation}
	The local density is evaluated either in the (denser) centroid of the potential ripples $\rho_{\textrm{centroid}}$, or averaged all over the volume of the emerging fragments $\widetilde{\rho}$, or, more significantly, corresponding to the matter contained in the volume of the potential ripples $\rho_{\textrm{well}}$.
	We deduce that, as expected from the calculation in stable nuclear matter discussed above, the isotopic width results underestimated with respect to the analytic prediction of Eq.~(\ref{eq:fluctuation_dissipation}). 
% variance which correspond to the correct Boltzmann-Langevin isovector fluctuation, even if in an open system. 
The difference is still acceptable due to the following two effects.
	First of all, fluctuations are built out of equilibrium: this implies that the collision rate is higher, generally leading to larger variances.
	Secondly, in open systems, particle evaporation may contribute in widening the isotopic spectra.

	The bottom row of Fig.~\ref{fig_isotopicdistributions} investigates the average isospin content measured in potential ripples in the corresponding early and late time intervals as a function of the $\rho_{\textrm{well}}$.
	The %IMF 
mass $A'$ grows with the density and the corresponding isospin content decreases, signing a process of isospin distillation.

\subsection{Extent of different clusterisation processes in heavy-ion collisions}
%
%	--- FIGURE isotopicdistributions
%
%
%

%
%	other similar processes
%
%		The processes discussed in this work, related to phase-space fluctuations and mechanical instabilities may dominate the dynamics of heavy-ion collisions at Fermi energies, between about 20 and 50~MeV per nucleon~\cite{Colonna2016}, or in spallation induced on heavy targets by light particles, like 1~GeV protons~\cite{Napolitani2015}: the involved phenomenology is very rich because these situations correspond to phase transitions and thresholds between very different mechanisms (fusion, fission, multifragmentation, neck emission...).

	The processes discussed in this work, related to phase-space fluctuations and mechanical instabilities, involve a rich phenomenology of phase transitions and thresholds between very different reaction mechanisms.
	They may therefore also present some similarities in their outcome with other rather different processes and, in some situations, combine with them.
	For instance, the onset of instabilities of Rayleigh type is a common process in macroscopic hydrodynamic systems, like classical fluids with a non-negligible surface tension~\cite{Ashgriz1990}, which has also been proposed as a possible additional scenario for nuclear multifragmentation in heavy-ion collisions~\cite{Moretto1992,Lukasik1997}, in some specific situations.
	Such process occurs in systems where a dilute core expands into a denser shell (Rayleigh-Taylor instability), or it acts on very deformed systems involving cohesional forces which respond to external perturbations (Plateau-Rayleigh instability).
	The system in such hydrodynamic scenario develops hole nucleation, evolving into a sponge-like or a filamented configurations which then relax into compact droplets.
	The mechanism is faster than ordinary fission and density variations of bound matter along the process do not need to be significant.

	On the other hand, the spinodal process described above can only occur if the system traverses a specific region of the equation of state, characterised by negative incompressibility where nucleation progresses from a dilute phase, letting blobs of larger density gradually emerge.
	From a microscopic point of view, it is rather associated to the nuclear liquid-gas phase transition and it requires a time comparable to the equilibration time of the system 
in reactions at Fermi energies. %\PN{ok?}
	We point out that, in simulations of heavy-ion collision, the BLOB approach is actually able to describe the interplay between spinodal processes and the above mentioned hydrodynamic effects~\cite{Napolitani2016}.

	Heavy-ion collisions and nuclear matter also involve processes of nuclear cluster formation, from light charge particles to heavier nuclear molecules, but those products emerge from an even different mechanisms~\cite{Typel2014}, which would require the explicit inclusion of additional correlations in the hierarchy of Eq.~(\ref{eq:BBGKY}).
	Light charged particles related to nuclear clustering have too small size, exceeding the ultraviolet cutoff of the dispersion relation, so that they can not belong to the unstable multipole modes which characterise spinodal fragmentation. 
	Solutions for an explicit treatment of cluster formation are proposed in refs.~\cite{Danielewicz1991,Kuhrts2001,Ono2016}.
	Connections between nuclear clustering and (spinodal) multifragmentation might be proposed, considering that multifragmentation might act on defining nuclear sources with rather complex shape from which clustered structures might eventually emerge. 
%which constitute the precursors of clustering, like isospin properties and configuration topology.

%isospin distillation, the amount of which is proportional to the repulsion of the symmetry term at sub-saturation densities.

%a liquid more symmetric than the global system

%
%	astrophysics
%
%	\PN{Speculations about astrophysical scenarios...about specificity of nuclear system with respect to stars, discussion around spinodal suppression}

%%%%%     %%  %%%  %%%     %%%     %%      %%      %%     %%%%%%%%%
%%%%%  %%%%%  %%%  %%  %%%  %%  %%  %%%  %%%%  %%%%%%  %%  %%%%%%%%
%%%%%  %%%%%       %%  %%%  %%  %%  %%%  %%%%    %%%%  %%  %%%%%%%%
%%%%%  %%%%%  %%%  %%       %%     %%%%  %%%%  %%%%%%     %%%%%%%%%
%%%%%     %%  %%%  %%  %%%  %%  %%%%%%%  %%%%      %%  %%  %%%%%%%%
\section{Conclusions  \label{sec_conclusion}}

	This work presents crucial steps to validate BL transport models applied to a fermionic system,
both in stable and mechanically unstable conditions, as far as the development of isoscalar and isovector
fluctuations at various densities is concerned. %in presence of mechanical instabilities and isovector fluctuations.
	In particular, the amplitude of fluctuations 
is investigated in relation with the corresponding properties of the nuclear effective interaction.
%are studied as a function of time in correspondence with the properties of the nuclear interaction.
	A transport approach constructed by requiring to satisfy the dispersion relation of 
mean-field unstable modes proves to be suited for the description of nuclear multifragmentation.
%	The transport approach constructed on the basis of these features proves to be able to describe the bulk dynamics of the spinodal process occurring in the mechanically unstable region of the equation of state.
%	Such a process was suggested as a possible description of nuclear multifragmentation.
%	Alternatively, other descriptions of multifragmentation were proposed relying on statistical approaches.
%	In this respect, it should be mentioned that the dynamical description presented in this work does not require any initial hypothesis on the thermodynamics of the system, like equilibration for instance, but the characteristic thermodynamic features of multifragmentation, like the occurrence of a nuclear liquid-gas phase transition, are obtained as a result of the dynamical description~\cite{Napolitani2013}, making this way the two approaches mutually consistent.

	In practice, even though technically demanding, the BLOB approach constitutes a conceptually  straightforward solution of the BLE in three dimensions.
	Through a simple renormalisation of the collision term, the Fermi statistics is in fact preserved for a long time and a correct isoscalar fluctuation amplitude is obtained independently of the ingredients of the numerical implementation (like the number of test particles), just associated with the collision rate and the growth time of the unstable modes. 
%	This feature is fundamental for describing fragment formation in a one-body description.
	Also the fluctuation variance of isovector observables are better treated than in conventional semiclassical approaches, even though the expected variance is still not achieved.  
%This is related to the variety of configurations, concerning 
%shape and extension of the nucleon wave
%packet, which occur 
	Indeed, in the implementation of the fluctuating collision integral, different configurations, varying 
in shape and extension, are possible to represent the nucleon wave packet.   
	This induces smearing effects on the fluctuation amplitude.  
	Moreover, if the collision rate is very low, the characteristic time scales associated with the construction of collisional two-body correlations are larger than the typical mean-field time scale and the fluctuations are damped by the propagation in the stable mean-field. 
	This explains why, even for equilibrated nuclear matter, the model yields a dependence of the isovector fluctuation amplitude on the nucleon-nucleon cross section. 
	It is also observed that, if two-body collisions are too rare, the numerical noise dominates the dynamics and one obtains a fluctuation variance reduced by a factor $1/\Ntest$ with respect to the expected value, as obtained in standard transport approaches where fluctuations are neglected~\cite{Bonasera1994}.        
	On the other hand, if the collision rate is large enough, the fluctuation amplitude does not depend on the number of test particles employed in the simulations. 
%	The BLOB approach allows for a full treatment of phase-space fluctuations which are continuously and spontaneously revived in time from collisional correlations, without any external action on the model.

	It is worth noting that the dynamical approach presented in this work does not imply any thermodynamic hypothesis (equilibration for instance) in the implementation of the fluctuation source, so that the characteristic thermodynamic features of multifragmentation, like the occurrence of a nuclear liquid-gas phase transition, are obtained as a result of the transport dynamics~\cite{Napolitani2013}: this finding makes the present dynamical description and alternative statistical approaches for multifragmentation mutually consistent.
	Finally, this approach can easily connect nuclear matter to heavy-ion collisions in the same framework.

\begin{acknowledgments}

	This project has received funding from the European Unions Horizon 2020 Research and Innovation
Programme under Grant Agreement No. 654002. 

	Research conducted in the scope of the International Associated Laboratory (LIA) COLL-AGAIN.

\end{acknowledgments}

%%%%%     %%  %%%  %%%     %%%     %%      %%      %%     %%%%%%%%%
%%%%%  %%%%%  %%%  %%  %%%  %%  %%  %%%  %%%%  %%%%%%  %%  %%%%%%%%
%%%%%  %%%%%       %%  %%%  %%  %%  %%%  %%%%    %%%%  %%  %%%%%%%%
%%%%%  %%%%%  %%%  %%       %%     %%%%  %%%%  %%%%%%     %%%%%%%%%
%%%%%     %%  %%%  %%  %%%  %%  %%%%%%%  %%%%      %%  %%  %%%%%%%%

\appendix
%\appendix*

%\section{Implementation of Boltzmann-Langevin approaches \label{sec_BLimplementation}}
\section{Exploiting fluctuations in BLOB: handling metrics and nucleon-nucleon collision statistics \label{sec_appendix_metrics}}

	In Sec.~\ref{sec_correlations_BLOB} the BLOB scheme, Eq.~(\ref{eq:BLOB}), is introduced to generate stochastic dynamical paths in phase space.
	The system is sampled through the usual test-particle method, often adopted for the numerical resolution of transport equations~\cite{Bertsch1988}
%{\bf [(5) add references]}
with the difference that, in the case of the BLOB implementation, the phase-space portions $A$ and $B$ involved in single two-body collisions are not two individual test particles but rather agglomerates of $\Ntest$ test particles of equal isospin, where $\Ntest$ is the number of test particles per nucleon used in the simulations. 
%to describe the mean field. 
	In a binary nucleon-nucleon collision, the initial states $A$ and $B$ are constructed by agglomeration around two phase-space sites, which are sorted at random, inside a phase-space cell of volume $h^3$, according to the method proposed in ref.~\cite{Napolitani2013} and further improved in ref.~\cite{Napolitani2015}.
	At successive intervals of time, by scanning all phase space in search of collisions, all test-particle agglomerates are redefined accordingly in $h^3$ cells, so as to continuously restore nucleon-nucleon correlations.
	Since test particles could be sorted again in new agglomerates to attempt new collisions in the same interval of time, the nucleon-nucleon cross 
section $\sigma_{N\!N}$ contained in the transition rate $W$ should be divided by $\Ntest$: 
\begin{equation}
	\sigma = \sigma_{N\!N} / \Ntest \;.
\label{eq:XSscaling}
\end{equation}

	Boltzmann-Langevin solutions where an ensemble of $\Ntest$ test particles are moved in one bunch and the nucleon-nucleon cross section is scaled by $\Ntest$ where already followed in the early approach by Bauer and Bertsch~\cite{Bauer1987}, or in more recent implementations~\cite{Mallik2015}.
	There is however a very fundamental difference: in the Bauer-and-Bertsch approach the Pauli-blocking term is not applied to the involved portions of phase space which are actually interested by the scattering at a given time $t$, as imposed by Eq.~(\ref{eq:Pauli_scaled}), but it is applied only to the centroids of
%to two single test particles belonging to 
the two colliding packets.
	Such approximation makes the Pauli blocking satisfied only approximately,
%in average over several collisions but does not respect the occupancy variance $f^{(n)}(1-f^{(n)})$, 
with the drawback of loosing the Fermi statistics~\cite{Chapelle1992}.
	In the direction of BLOB, to prevent the above problem, a first practical solution %in one dimension 
was proposed in Ref.~\cite{Rizzo2008}.
%{\bf [(6) comments on the work by Malik]}

\begin{figure}[t!]
\begin{center}
	\includegraphics[angle=0, width=1\columnwidth]{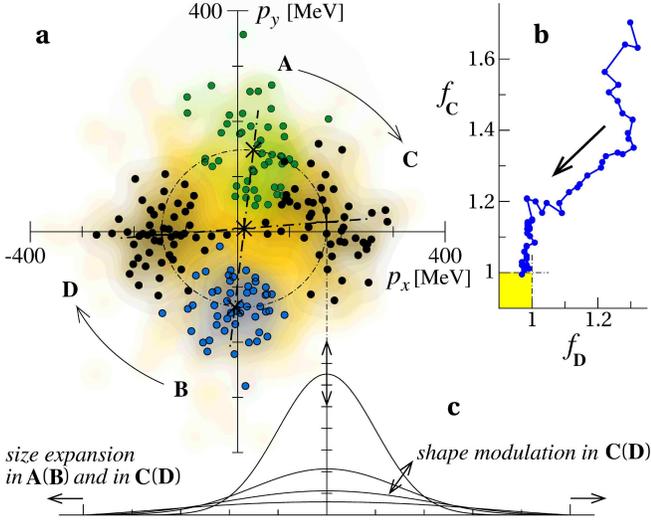}
\end{center}\caption{
(a) Schematic representation of the definition of test-particle agglomerates in their initial ($A$,$B$) and final ($C$,$D$) states in momentum space in a $h^3$ volume.
(b) Convergence of a binary nucleon-nucleon collision configuration towards a situation where Pauli blocking is strictly satisfied. The path in the plane 
determined by the occupancy numbers, $f_C$ and $f_D$, of the final states
collects the sequence of modulations in phase-space of the test-particle clouds where the occupancy of the destination regions is iteratively optimised.
(c) Examples of possible shape modulations for the packets associated with
nucleons in the final state.
\label{fig_modulation} }
\end{figure}

	Moreover, in BLOB, special attention is paid to the metrics when defining the test-particle agglomeration: the agglomerates are searched requiring that they are the most compact configurations in the phase space metrics which does neither violate Pauli blocking in the initial and in the final states, nor energy conservation in the scattering.
%	The localisation in momentum space makes the collisions more effective in agitating the phase space, and the localisation in coordinate space is needed to describe hydrodynamical effects like the flow dynamics.
	For this purpose, when a collision is successful, its configuration is further optimised by modifying the shape and the width of the initial and final states~\cite{Napolitani2012}.
	Fig.~\ref{fig_modulation} illustrates the paths of a collision configuration which by a procedure of successive modulations is brought to a situation which respects Pauli blocking strictly. 
	If such modulation procedure results unsuccessful, the collision is rejected.
	The rate of rejections due to unsuccessful modulation of the collision configuration is close to zero in open systems (heavy-ion reactions) so that the correlation between attempted and effective collision number is identical if a UU or a BLOB collision term is applied, %to the same mean-field, 
provided that the same nucleon-nucleon cross section is used.
	On the other hand, in uniform nuclear matter 
at equiliubrium, where only nucleons close to the Fermi surface can be involved
in two-body collisions,
%in conditions of high density (close to the saturation density and above) 
the occurrence of such rejections becomes not negligible when the
temperature $T$ considered is very low compared to the Fermi momentum.
	In this case, the exact correspondence between attempted and effective collision rates in BUU (or SMF) and BLOB is lost.
%{\bf [(7) il problema non e' l'alta densita', ma il fatto che la temperatura e' piccola rispetto all'impulso di Fermi. Se aumentassimo la temperatura, le rejections diminuirebbero, anche ad alta densita']}

%
%	--- FIGURE modulation
%

	A remarkable advantage of the renormalised form of the residual contribution in Eq.~(\ref{eq:BLOB}) is to connect directly the fluctuation variance to the physical  properties of the system, regardless the test-particle number.
	Such aspect has a general relevance because it makes the dynamics independent from many aspects of the numerical implementation.
	The dependence on $\Ntest$ persists on the other hand in the mean-field representation, therefore when the physical fluctuation amplitude is small, the global fluctuation phenomenology may suffer from noise effects produced by the use of a finite number of test particles in the numerical implementation of the transport equation.
	This remark should be kept in mind for the study of fluctuations of relatively small amplitude, like isovector fluctuations, as discussed in Sec.~\ref{sec_ivfluctuations_noise}.
%	The interplay between physical and numerical fluctuations will be discussed in more details in the following.

\section{Model parameters for nuclear matter \label{sec_appendix_parameters}}

%
%	--- FIGURE Fermi-Dirac distribution
%
\begin{figure}[b!]\begin{center}
	\includegraphics[angle=0, width=.8\columnwidth]{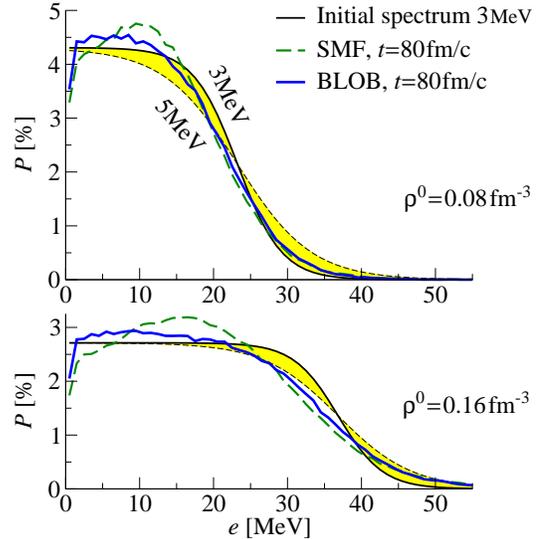}
\end{center}\caption
{
	Energy distribution for $\rho^0=0.08$ (top) and $0.16$~fm$^{-3}$ (bottom), as obtained in SMF (dashed line) and BLOB (full line) calculations at 
 $t=80$~fm/$c$.
	The coloured band connects the Fermi-Dirac distributions 
corresponding to $T=3$ and $T=5$~MeV. 
%the first value corresponding to the initial conditions.
%	Spectra evaluated at $t=80$~fm/$c$, through the SMF and BLOB approaches are overlapped.
%	An asy-stiff parameterisation is used.
}
\label{fig_Fermi_Dirac}
\end{figure}
%
%

%	In the following BLOB and SMF are used to simulate aspects of fluctuation phenomenology in nuclear matter.
	In this work, for comparison purposes, both BLOB and SMF models are prepared as relying on a strictly identical implementation of the mean field, so that they differ only for the residual contribution.
	A simplified Skyrme-like (\textit{SKM}$^*$) effective interaction~\cite{Guarnera1996,Baran2005}, where momentum-dependent terms are omitted, is employed in the propagation of the one-body distribution function, corresponding to the following definition of the potential energy per nucleon:
\begin{equation}
	\frac{E_{\textrm{pot}}}{A}(\rho) = 
		\frac{A}{2}u
		+\frac{B}{\sigma+1}u^\sigma
		+\frac{C_{\textrm{surf}}}{2\rho}(\nabla\rho)^2
		+\frac{1}{2}C_{\textrm{sym}}(\rho)u\beta^2 ,
\label{eq:pot}
\end{equation}
with $u=\rho/\rho_{\textrm{sat}}$, being $\rho_{\textrm{sat}}$ the saturation density %\PN{=0.145} 
and $\beta=(\rho_n-\rho_p)/\rho$.
	This parameterization, with $A\!=\!-356$ MeV, $B\!=\!303$ MeV and $\sigma\!\!=\!7/6$, corresponds to a soft isoscalar equation of state with a compressibility modulus $K\!=\!200$~MeV.
	An additional term as a function of the density-gradient introduces a finite range of the nuclear interaction and accounts for some contribution from the zero-point motion of nucleons~\cite{Guarnera1996}.
	$C_{\textrm{surf}}$ is related to various properties of the interaction range: the surface energy of ground-state nuclei (the best fit imposing a value of $-6/\rho_{\textrm{sat}}$ MeV fm$^5$), the surface tension (light-fragment emission, in comparison to available data is better described for a smaller range given by $-7/\rho_{\textrm{sat}}$ MeV fm$^5$ in BLOB), and the ultraviolet cutoff in the dispersion relation for wavelengths in the spinodal instability (larger spectrum for a smaller range) \cite{Ayik1995}.
%	A value of $C_{\textrm{surf}}\!=\!-7/\rho_0$ MeV fm$^5$ is chosen in this work.
%
%-------------------------------------------------------------
%	\PN{See also Jacquot PhD pag 52 (parameter a0)!!!}
%
%	Initial cond. Prepanda (Gaussian functions):
%	FWHM:    gr = 1.4446
%	Sigma:   sigr=gr/(2*SQRT(2LOG(2))) = 0.6135
%
%	Twingo (triangular functions):
%	Sigma:   sigrsqrt2 = grsqrt2/(2*SQRT(2LOG(2))) = 0.868
%	Corr.:   surf = 7.00
%-------------------------------------------------------------
%
	In this work, a value of $C_{\textrm{surf}}\!=\!-7/\rho_0$ MeV fm$^5$ is chosen for the surface term.
	A linear (asy-stiff) density dependence of the potential part
of the symmetry energy coefficient, $E_{\textrm{sym}}^{\textrm{pot}}$, is obtained by setting $C_{\textrm{sym}}(\rho)\!=\!{\textrm{constant}}\!=\! 32$~MeV and a quadratic-like (asy-soft) dependence is obtained for $C_{\textrm{sym}}(\rho)\!=\! \rho_{\textrm{sat}}(482-1638\rho)$ MeV~\cite{Colonna2014}.

	$\Ntest=40$ test particles per nucleon are employed if not otherwise specified.
	In this work the collision term involves an isospin- and energy-dependent free nucleon-nucleon cross section with an upper cutoff at $\sigma_{\textrm{NN}}=50$~mb \cite{Napolitani2013}.
	In some cases, when indicated, these prescriptions may have been modified.
	In the SMF approach we adopt the quite short time interval of 2 fm$/c$ to inject fluctuations.
	In unstable conditions, like the spinodal region studied in Sec.~\ref{sec_isfluctuations}, the choice of a short time interval as compared to the typical growth time of unstable modes leads to convergent results on isoscalar fluctuations. 
	We note that the growth time of unstable modes amounts to about 30 fm$/c$., quite independently of the density conditions (see Sec.~\ref{sec_disprel}).
	Concerning isovector fluctuations, they are only induced by the finite number of test particles. 
	Indeed, in the SMF model there are no explicit fluctuation terms injected in the isovector channel.

	To simulate nuclear-matter, we prepare the system in a cubic periodic box of edge size $L=39$~fm, and we subdivide it in a lattice of cubic cells of edge size $l$ where we calculate density variances.
	For the sake of simplicity, we consider symmetric nuclear matter, i.e. with
equal number of neutrons and protons.
%	We initially define the system imposing a uniform-matter effective field $U^0(\vecp)$ which does not depend on configuration space; correspondingly we define a uniform-matter density $\rho^0(\vecp)$ and distribution function $f^0(\vecp)$ depending only on momentum space.
	We initially define the system by imposing a uniform-matter effective field $U^0(\rho)$ whose amplitude only depends on the density considered,
%momentum magnitude, 
%{\bf [(9) perche' il campo medio dipenderebbe dall'impulso ???]}
and a corresponding effective Hamiltonian $\epsilon(p) = h^0(p) = p^2/(2m) + U^0(\rho)$.
	Accordingly, the phase-space distribution function $f^0(\vecp)= \{1+\exp[\epsilon(p)-\mu]/T\}^{-1}$, not depending on configuration space (because the
system is homogeneous), is the Fermi-Dirac equilibrium distribution at the 
temperature $T$ and chemical potential $\mu$.

%%%%%%%%%%%%%%%%%%%%%%%%%%%%%%%%

%	To prepare a transport calculation, the system is initialised with a Fermi-Dirac distribution at a temperature $T=3$~MeV; it is then sampled for several values of $\rho^0$ and the potential, restricted to the only isovector contribution, is tested for a stiff and a soft density dependence of the symmetry energy (see appendix~\ref{sec_appendix_parameters}).

%	To prepare a transport calculation, the system is sampled for several values of $\rho^0$ and the potential, restricted to the only isovector contribution, is tested for a stiff and a soft density dependence of the symmetry energy (see appendix~\ref{sec_appendix_parameters}).

	The system is initialised with a Fermi-Dirac distribution at a temperature $T=3$~MeV. 
	As shown in Fig.~\ref{fig_Fermi_Dirac}, both SMF and BLOB transport dynamics succeed to preserve the initial distribution quite efficiently as a function of time, even though a flattening of the spectrum around an effective equilibrium temperature
$T_{\textrm{eq}}$ should be accounted for,
%towards a larger equilibrium temperature, 
due to the fact that the Fermi statistics is not perfectly preserved.
	This temperature modification depends on the parameters of the calculation and is larger for larger densities.

\end{document}